\documentclass{aa}  

\usepackage{natbib}
\usepackage{diagbox}
\usepackage{slashbox}
\usepackage{multirow}
\usepackage{booktabs}
\usepackage{graphicx,subfigure}
\usepackage{txfonts}
\usepackage{hyperref}
\usepackage{appendix}
\newcommand{\urlNewWindow}[1]{\href[pdfnewwindow=true]{#1}{\nolinkurl{#1}}}
\usepackage{color}
\newcommand{\G}{\mathcal{G}}
\newcommand{\Msun}{M_\odot}

\newcommand{\Ms}{M_{\star}}
\newcommand{\Rs}{R_{\star}}
\newcommand{\Jconv}{J_{\rm conv}}
\newcommand{\Iconv}{I_{\rm conv}}
\newcommand{\Osconv}{\Omega_{\rm conv}}
\newcommand{\Jrad}{J_{\rm rad}}
\newcommand{\Irad}{I_{\rm rad}}
\newcommand{\Osrad}{\Omega_{\rm rad}}
\newcommand{\Is}{I_\star}
\newcommand{\Mjup}{M_{\rm jup}}

\newcommand{\Ts}{T_\star}
\newcommand{\Mp}{M_{\rm p}}

\newcommand{\dd}{\mathrm{d}}
\newcommand{\Os}{\Omega_{\star}}

\usepackage{epstopdf}
\usepackage[]{sidecap}

\begin{document} 

  \title{Planetary tidal interactions and the rotational evolution of low-mass stars}
  \subtitle{The Pleiades' anomaly}
  \author{F. Gallet\inst{1,4} \and E. Bolmont\inst{2,3}\and J. Bouvier\inst{4}, S. Mathis\inst{2,3} \and C. Charbonnel\inst{1,5}
          }

\offprints{F. Gallet,\\ email: florian.gallet1@univ-grenoble-alpes.fr}

  \institute{$^1$ Department of Astronomy, University of Geneva, Chemin des Maillettes 51, 1290 Versoix, Switzerland \\
$^2$ IRFU, CEA, Universit\'e  Paris-Saclay, F-91191 Gif-sur-Yvette, France \\
$^3$ Universit\'e  Paris Diderot, AIM, Sorbonne Paris Cit\'e , CEA, CNRS, F-91191 Gif-sur-Yvette, France \\
$^4$ Univ. Grenoble Alpes, CNRS, IPAG, 38000 Grenoble, France \\
$^5$ IRAP, UMR 5277, CNRS and Universit\'e de Toulouse, 14, av. E. Belin, F-31400 Toulouse, France \\
             }

  \date{Received -- ; Accepted -- }

  \abstract
  {The surface angular velocity evolution of low-mass stars is now globally understood and the main physical mechanisms involved in it are observationally quite constrained. However, while the general behaviour of these mechanisms is grasped, their theoretical description is still under ongoing work. This is the case for instance about the description of the physical process that extracts angular momentum from the radiative core, which could be described by several theoretical candidates. Additionally, recent observations showed anomalies in the rotation period distribution of open cluster, main sequence, early K-type stars that cannot be reproduced by current angular momentum evolution models.}
   {In this work, we study the parameter space of star-planet system's configurations to investigate if including the tidal star-planet interaction in angular momentum evolution models could reproduce the anomalies of this rotation period distribution.} 
   {To study this effect, we use a parametric angular momentum evolution model that allows for core-envelope decoupling and angular momentum extraction by magnetized stellar wind that we coupled to an orbital evolution code where we take into account the torque due to the tides raised on the star by the planet. We explore different stellar and planetary configurations (stellar mass from 0.5 to 1.0 $\rm M_{\odot}$ and planetary mass from 10 $\rm M_{\oplus}$ to 13 $\rm M_{\rm jup}$) to study their effect on the planetary orbital and stellar rotational evolution.}
   {The stellar angular momentum is the most impacted by the star-planet interaction when the planet is engulfed during the early main sequence phase. Thus, if a close-in Jupiter-mass planet is initially located at around 50\% of the stellar corotation radius, a kink in the rotational period distribution opens around late and early K-type stars during the early main sequence phase.} 
   {Tidal star-planet interactions can create a kink in the rotation period distribution of low-mass stars, which could possibly account for unexpected scatter seen in the rotational period distribution of young stellar clusters.}
 

\maketitle


\sidecaptionvpos{figure}{c}

\section{Introduction}

The angular momentum evolution of young low-mass stars has been investigated for several decades \citep[e.g.][]{WD67,Sku72,Kawaler88,Keppens95,Bouvier08,RM12}. The associated theoretical efforts led to a better understanding of the main physical mechanisms at work in this evolution \citep[star-disk interaction, magnetic braking, and internal redistribution of angular momentum, see e.g.][and references therein]{Bouvier14}. Strong theoretical constraints have been added to the processes that drive angular momentum transport in stellar interiors \citep[e.g.][]{Amard15} and the extraction of angular momentum by magnetized stellar winds \citep{matt15,Reville15,See17}. The physics that controls the angular velocity evolution of low-mass stars from the pre-main sequence (hereafter PMS) up to the end of the main sequence (hereafter MS) is thus now relatively well understood \citep[see e.g.][]{MP05a,GB15,Somers15,Lanzafame15,Amard15,Johnstone15,SA17}. Despite the uncertainties about the correct physical description to use to describe the mechanisms involved in the internal transport of angular momentum between the radiative core and the convective envelope, current models grasp the main trends of stellar rotational evolution.

In parallel to these theoretical developments, numerous exoplanets have been detected since 1995 \citep{MQ95} and now reach a number of confirmed objects between 2950 and 3786 (June 5 2018, see exoplanets.org and exoplanet.eu). These exoplanets can be found in a wide range of star-planet configurations that encompass a large distribution of planetary masses ranging from $10^{-4}$ to 100 $\rm M_{\rm Jup}$, orbital periods from $10^{-1}$ to $10^{5}$ days, and a (sub)stellar mass ranging from $2\times10^{-2}$ to 4 $\rm M_{\odot}$.  Nevertheless, most angular momentum evolution models mainly focus on isolated stars thus neglecting the possible impact of a planetary companion on the rotational evolution of the central star. However, the presence of close-in planets ought to be included in such numerical codes as pointed out by \citet{Bolmont12}, \citet{Zhang14}, \citet{Lanza16}, \citet{Privitera2016}, and \citet{Rao18}, who show the strong impact of the stellar rotational history on the orbital evolution of massive close-in planets, and vice-versa.

During the last decade, thanks to the inauguration of the Kepler satellite and most recently because of Kepler's second life mission K2, we have entered a new era of improved rotational period measurements that allows for advanced astrophysical quests. Indeed, the precision of measured stellar surface rotation periods (through photometric variation induced by the presence of magnetic stellar spots) is now good enough to detect specific features in the rotation period distribution of open clusters. This is for instance the case for the rotation period distribution of the Pleiades cluster (a 120 Myr old MS open cluster located at about 140 pc from the Earth) that has been analysed by \citet{Rebull16} and \citet{Stauffer16} using Kepler-K2 \citep{K2}. In this cluster, they found K-type stars 
with a faster rotation rate than expected from current theoretical angular velocity evolution tracks (hence producing a ``kink'' in the rotational distribution). These ``classical'' rotational tracks are produced by numerical models that only invoke star-disk interaction, angular momentum extraction by stellar wind, and internal redistribution of angular momentum within the stellar interior; but this anomaly could potentially result from the presence of an exoplanet that may affect, through tidal interaction, the surface rotation rate of its host star. Indeed, 
\citet[][see also \citealt{Bolmont16} and \citealt{Gallet17b}]{Mathis15} showed that the tidal dissipation inside the star is maximum, during the early-MS phase, for early K-type stars due to a specific configuration of their internal structure.

In the theoretical framework, tides in stars can be described by two components: the equilibrium tide, which corresponds to the large-scale hydrostatic adjustment induced by the gravitational interaction between the star and its companion \citep[of stellar or planetary nature, see][]{Zahn66,Remus2012,Ogilvie13} and which is made up of a large-scale non-wavelike/equilibrium flow; and the dynamical tides, which correspond to the dissipation of tidal inertial waves (mechanical waves that are generated inside rotating fluid bodies) due to the turbulent friction in convective regions \citep{Ogilvie07,Ivanov13} and to thermal diffusion and breaking mechanisms acting on gravito-inertial waves (gravity waves  influenced by the effect of rotation through the Coriolis acceleration) in radiative regions \citep[e.g.][]{Zahn75,Terquem98,Barker10}. 
The dissipation of the dynamical tides thus strongly depends on the internal structure of the star \citep[see e.g.][]{Chernov13,Ogilvie14,Mathis15,Gallet17b}.

Most of the studies dedicated to tidal star-planet interactions often assume solid body rotation for the whole star \citep[e.g.][]{Bolmont12,Nordhaus13,Bolmont16}. However, recent works have allowed for stellar core-envelope decoupling so as to investigate the impact of the presence of a massive planet on the surface rotation of the star. From the literature, \citet{Zhang14} used constant tidal dissipation efficiencies along the stellar evolution; \citet{Penev14} and \citet{Penev18} included core-envelope decoupling so as to add constraints on the evolution of the tidal dissipation; and \citet{Privitera2016} focused on star-planet interaction during the red giant phase. 

In this article, and in complement to the work of \citet{Bolmont16}, we investigate the impact of tidal dissipation evolution (controlled by the internal stellar structure during the PMS and by the surface rotation rate of the star during the MS) on the evolution of the rotation rate of the host star using a two-zone rotational model that allows for core-envelope decoupling. We explore the parameter space of star-planet systems, considering stellar mass, initial parameters (rotation, disk lifetime, and coupling timescale), planetary mass, and initial orbital distance to map the impact of star-planet interaction on the rotational evolution of low-mass stars. {{\citet{Rao18} also recently studied the impact of the equilibrium and dynamical tides on the orbital evolution of massive close-in planets. In their work they focused on the initial conditions that affect the planetary survivability around stars more massive than 1.0 $\rm M_{\odot}$ , while we are more interested in how the surface rotation rate of the host star is modified by the star-planet tidal interaction. These two works are thus very complementary.}}

This paper is structured as follows. The numerical model used in this work is described in Sect. \ref{model}. In Sect. \ref{rotevol} we investigate the rotational evolution of low-mass stars in the presence of a close-in planet, and how it is impacted by the main star-planet parameters. We first study the case of a solar mass star in Sect. \ref{referencecase}, and we study the evolution of a initial rotational distribution orbiting an early-K 0.8 $\rm M_{\odot}$ type star in Sect. \ref{distribinit}. Finally we generalize these results to a broader mass range in Sect. \ref{stellarmass}. We finally compare the results of our simulations to the Pleiades data in Sect. \ref{explomass} and conclude in Sect. \ref{conclusion}.

\section{Model}
\label{model}

The numerical model used in this work combines the rotational evolution model of \citet{GB13,GB15} and the modified orbital evolution model of \citet[][see \citealt{Bolmont16}]{Bolmont12}. The dissipation of the dynamical tide inside the star is treated in the convective envelope as in \citet[][who followed \citealt{Ogilvie13} and \citealt{Mathis15}]{Gallet17b} and the stellar structure is from the stellar evolution code STAREVOL \citep[see][and references therein]{Amard15}. We developed this code that combines these two numerical approaches so as to study, in a more realistic way through the addition of the decoupling between the radiative core and the convective envelope, the impact of the star-planet interaction on the surface rotation rate of low-mass stars. This code is specifically designed for stars between 0.3 and 1.2 $\rm M_{\odot}$ and for binary star-planet systems consisting of one central object and one planet. 

In the following we recall the specificities of each part of our combined numerical model.

\subsection{Stellar evolution}

This work is based on a grid of non-rotating stellar models we computed with the code STAREVOL \citep[see e.g.][]{Amard15} for a range of initial masses between 0.5 and 1.0~M$_{\odot}$ at solar metallicity (Z = 0.0134; \citealt{AsplundGrevesse2009}).
The {references for the} basic input microphysics (equation of state, nuclear reactions, and opacities) can be found in \cite{Amard15} and in \cite{Lagardeetal12}. The initial {helium} abundance and mixing length parameter are calibrated without {atomic}  diffusion to reproduce a non-rotating Sun with respect to the solar mixture of \cite{AsplundGrevesse2009} with a $10^{-5}$ precision for the luminosity and the radius at the age of the Sun. The corresponding mixing length parameter and initial helium abundance are $\alpha_{\rm{MLT}} = 1.6267$ and $Y = 0.2689$.
The stellar structure provided by the STAREVOL code is the foundation of our model. It is used on the one hand to follow the evolution of the stellar rotation rate and on the other hand to estimate the tidal dissipation inside the stellar interior.

\subsection{Tidal dissipation}
\label{tidaldissip}

Following the studies of \citet{Bolmont16}, \citet{Gallet17b}, and \citet{Bolmont17}, we compute the tidal evolution of planets using a model improved with respect to the classical orbital evolution models, which only take into account the equilibrium tide \citep[e.g.][]{Mignard1979, Hut1981, Bolmont11, Bolmont12, Bolmont2015}.
The equilibrium tide consists of a large-scale flow driven by the hydrostatic adjustment of the body due to the perturbing gravitational potential of the planet. For rotating bodies, inertial waves, which are driven by the Coriolis acceleration, can be excited for a certain range of excitation frequencies ($\omega \in [-2\Os, 2\Os]$, where $\omega \equiv 2\left( n-\Omega_{\star} \right)$ is the tidal frequency in the case of circular coplanar systems, $n$ is the mean orbital motion, and $\Omega_{\star}$ is the surface rotation rate of the star) in the convective envelope of low-mass stars. Assuming a two-layer star in solid body rotation, \citet{Ogilvie13} derived the frequency-averaged tidal dissipation induced by the dynamical tide, constituted by tidal inertial waves, in the convective envelope. 
In the case of a coplanar star-planet system in which the orbit of the planet is circular, this dissipation is given by
\begin{eqnarray}
\label{dissipequa}
<\mathcal{D}>_{\omega} =& \int^{+\infty}_{-\infty} \rm{Im}\left[k_2^2(\omega)\right] \displaystyle \frac{\dd\omega}{\omega} = \displaystyle \frac{100\pi}{63}\epsilon^2 \left( \displaystyle \frac{\alpha^5}{1-\alpha^5} \right)  \left( 1-\gamma \right)^2 \\ 
\times&  \left( 1-\alpha \right)^4  \left( 1 +2\alpha+3\alpha^3 + \displaystyle \frac{3}{2}\alpha^3 \right)^2 \left[ 1+  \left( \displaystyle \frac{1-\gamma}{\gamma} \right) \alpha^3  \right] \nonumber \\ 
\times& \left[   1 + \displaystyle \frac{3}{2}\gamma + \displaystyle \frac{5}{2\gamma}  \left( 1 + \displaystyle \frac{1}{2}\gamma - \displaystyle \frac{3}{2}\gamma^2 \right)  \alpha^3 - \displaystyle \frac{9}{4} \left( 1-\gamma \right)\alpha^5   \right]^{-2} \nonumber,
\end{eqnarray}
with 
\begin{eqnarray}
\alpha = \displaystyle \frac{R_{\rm{rad}}}{R_{\star}}, {\rm~} \beta=\displaystyle \frac{M_{\rm{rad}}}{M_{\star}}, {\rm~} \gamma =\displaystyle \frac{\alpha^3(1-\beta)}{\beta(1-\alpha^3)} < 1, \epsilon = \displaystyle \frac{\Os}{\Omega_{c}},
\end{eqnarray}
where $\Omega_{c}$ is the critical angular velocity of the star, $k_2^2$ is the Love number of degree 2 (corresponding to the quadrupolar component, $l=2$ and $\rm M=2$, of the time-dependent tidal potential proportional to the spherical harmonic $Y_l^m$), $R_{\rm{rad}}$ and $\rm M_{\rm{rad}}$ are the radius and mass of the radiative core, respectively, and $R_{\star}$ and $\rm M_{\star}$ are the radius and mass of the whole star, respectively. When present, the convective envelope surrounds the radiative core and both regions are assumed to be homogeneous with respective average densities $\rho_{\rm c}$ and $\rho_{\rm e}$. The ratio between the perturbation of the gravitational potential induced by the presence of the planetary companion and the tidal potential evaluated at the stellar surface is given by $k_2^2$. Its imaginary component $\rm{Im}\left[k_2^2(\omega)\right]$ is a direct estimation of the tidal dissipation. 
This formalism is very {convenient} because it allows us to take into account the dependence of tidal dissipation on stellar structure and rotation along their evolution, filtering out the complex frequency dependence of the dissipation of tidal inertial waves \citep{Ogilvie07}. This constitutes a first step for the study of the secular tidal evolution of star-planet systems \citep[we refer the reader to][for detailed discussion]{Mathis15,Bolmont16,Gallet17}.

\begin{figure}
\begin{center}
   \includegraphics[width=\linewidth]{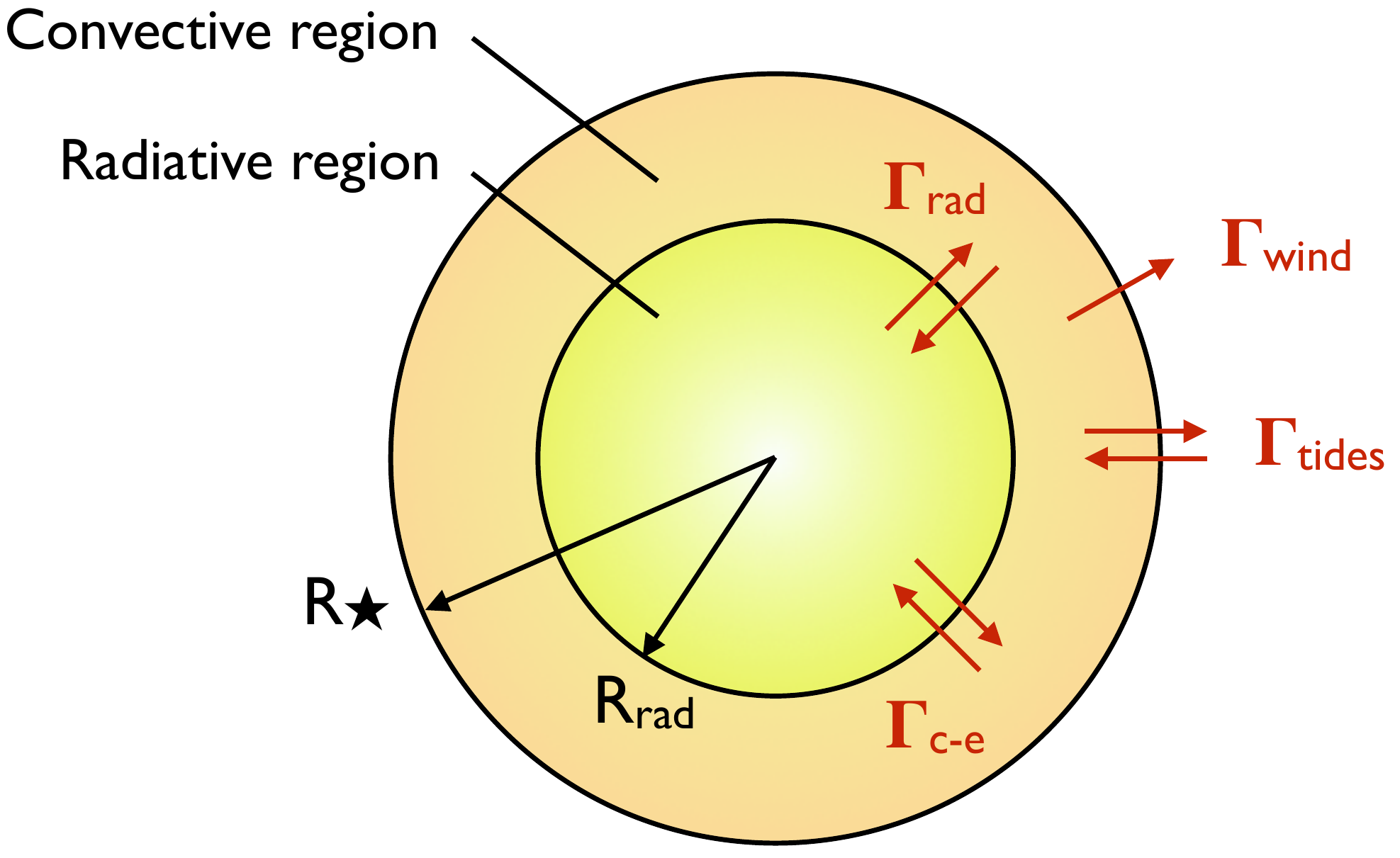}
   \caption{Sketch of the stellar structure studied here. We consider a two-layered star: a radiative core of rotation rate $\Omega_{\rm rad}$ and a convective envelope of rotation rate $\Omega_{\rm conv}$. Both are assumed to rotate as a solid body with different rotation rates. In red we indicate the different processes included in the model. The convective envelope is subjected to external torques (stellar wind and tidal torques), as well as torques due to the radiative core development and coupling with the convective envelope.}
\label{schema}%
\end{center}
\end{figure}
In order to compute the orbital evolution of close-in planets, we use the model introduced in \citet{Bolmont16}. The evolution of the semi-major axis $a$ of a planet on a circular orbit is given by \citep{Hansen10,Leconte2010,Bolmont11,Bolmont12} 
\begin{equation}\label{Hansena}
\displaystyle \frac{1}{a}\displaystyle \frac{\dd a}{\dd t} = - \displaystyle \frac{1}{\Ts}\Big[1-\displaystyle \frac{\Os}{n}\Big],
\end{equation}
where $\Ts$ is the evolution timescale given in Eq.~(\ref{Ts}).
It depends on the stellar equivalent structural quality factor $\overline{Q_s'}$, which can be expressed in terms of $<\mathcal{D}>_{\omega}$ as
\begin{equation}\label{Qs_Dw}
<\mathcal{D}>_{\omega} = \displaystyle \frac{3}{2 \overline{Q'}} = \epsilon^2 \displaystyle \frac{3}{2 \overline{Q_s'}},
\end{equation}
where $\overline{Q'}$ is the equivalent modified tidal quality factor as defined in \citet{Ogilvie07} and \citet{Mathis15b}. The tabulated values of $\overline{Q'}$ and $\overline{Q_s'}$ used in this study can be found \href{https://obswww.unige.ch/Recherche/evol/starevol/Galletetal17.php}{here.}\footnote{https://obswww.unige.ch/Recherche/evol/starevol/Galletetal17.php.}

The tidal torque exerted by the planet on the convective envelope of the star is given by \citep[e.g.][]{Bolmont16, Gallet17b}
\begin{eqnarray} 
\label{torque_tid}
\Gamma_{\rm tide}  & = \displaystyle \frac{h}{2\Ts}\left[1-\displaystyle \frac{\Os}{n}\right],
\end{eqnarray}
with $h$ the orbital angular momentum of the planet, $n$ its orbital frequency, and $\Ts$ an evolution timescale given by
\begin{equation}
\label{Ts}
\Ts = \displaystyle \frac{2}{9}\displaystyle \frac{\Ms}{\Mp(\Mp+\Ms)}\displaystyle \frac{a^8}{\Rs^{5}}\displaystyle \frac{\overline{Q_s'}}{\hat{\epsilon}^2}\displaystyle \frac{|n-\Os|}{\G},
\end{equation}
which relies on the semi-major axis $a$ of the planet, the mass ($\Ms$) and radius ($\Rs$) of the star, the mass $\Mp$ of the planet, the stellar equivalent structural tidal quality factor $\overline{Q_s'}$ \citep[see][and references therein]{Gallet17b,Bolmont17}, and $\hat{\epsilon} = \left(  \Omega_{\star} / \sqrt{\G M_{\odot}/R_{\odot}^3}  \right) \equiv \Omega_{\star}/\Omega_{\odot,c}$ with $\Omega_{\odot,c}$ the critical angular velocity of the Sun. The stellar equivalent structural tidal quality factor $\overline{Q_s'}$ strongly and primarily depends on the evolution of the stellar internal structure (i.e. the relative mass and size of the radiative core compared to the whole star). It evolves over five orders of magnitude during the stellar evolution \citep[see][]{Gallet17b}. For the equilibrium tide, we use a constant value for the dimensionless dissipation factor $\bar{\sigma}_{\star}$ \citep[which is linked to $1/\overline{Q_s'}$, see][]{Bolmont16}\footnote{$\bar{\sigma}_{\star}  = k_2^2  \displaystyle \sqrt{\frac{M_{\odot} R_{\odot}^7  }{\mathcal{G}}} \frac{2 \mathcal{G}}{ R_\star^5}  \frac{3\hat{\epsilon}^2 }{4 \overline{Q_s'}|n-\Omega_\star|} $}, which only depends on the stellar mass and is calibrated on the observation \citep[see][]{Hansen12}.

In this work, the tidal dissipation (and thus the tidal torque) in the radiative core is not included \citep{Zahn75,Goodman98,Terquem98,Barker10,Barker11,Ivanov13}. This additional torque could have a strong influence on the orbital evolution of close-in planets (it might make them fall significantly faster into the stars) and thus changes the way the surface rotation is impacted by the tidal interaction. The internal differential rotation inside the star has also a clear effect on the rotational evolution of the convective envelope \citep[especially during the MS phase, see][]{GB13,GB15}. A change of rotation rate due to the tidal torque applied in the radiative core could then affect the rotation of the envelope. We also do not include the dissipation inside the planet itself. {In this work, we do not consider massive close-in planets that were formed with the high-eccentricity scenario. This hypothesis is consistent with the fact that we neglect planetary tides that arise if the orbit is eccentric or if the rotation is not synchronized with zero obliquity. We also consider that the planet is alone (or at least that it does not feel any gravitational pull from another planet), which is also consistent with the disk migration of a single massive planet. Indeed, this latter should have destabilized its close environment along its evolution.}

{Finally, in the case of a planetary engulfment we assume that the whole angular momentum of the planet is instantaneously transferred to the star which end up in the rapid increase of the stellar surface angular velocity.}

\subsection{Stellar rotational evolution}

We model the evolution of the rotation of the central star using the formalism of \citet{GB15} in which the star is assumed to be composed of two parts: a radiative core surrounded by a convective envelope (see Fig. \ref{schema}). As in the averaged-dissipation tidal model, both regions are assumed to rotate as a solid body with different rotation rates. 
The stellar angular momentum is described by $J_{\star} = I_\star \Os$, where $\Is \propto M_{\star}R_{\star}^2$ is the moment of inertia. The total stellar angular momentum is thus defined by $J_{\star} = J_{\rm rad} +  J_{\rm conv}$, where $J_{\rm rad}$ is the angular momentum of the radiative region, with $R_{\rm rad}$ its radius and $\Osrad$ its rotation rate, and $J_{\rm conv}$ is the angular momentum of the convective region and $\Osconv$ its rotation rate.
The total angular momentum evolution rate is given by
\begin{eqnarray} 
\label{dJdt}
  \displaystyle \frac{\dd J_{\star}}{\dd t} = \displaystyle \frac{\dd \Is}{\dd t}\Os + \Is \displaystyle \frac{\dd \Os}{\dd t} = \Gamma_{\rm all},
\end{eqnarray}
where $\Gamma_{\rm all}$ is the sum of the external torques; here we account for the stellar wind $\Gamma_{\rm wind}$ and the tidal torque $\Gamma_{\rm tide}$; we neglect other external torques such as the accretion torque $\Gamma_{\rm acc}$ and the star-disk interaction torque $\Gamma_{\rm disk}$ since our simulations start after proto-planetary disk dissipation. The magnetic star-planet interactions $\Gamma_{\rm mag}$ are also not included in this work (see \citealt{Strugarek17} for detailed discussion about the range of application of the magnetic star-planet torque).
While the wind always acts to remove angular momentum ($\Gamma_{\rm wind} < 0$), the tidal interaction can either spin up or spin down the star depending on whether the planet is located inside or outside of the corotation radius \citep[e.g.][]{Bolmont16}
\begin{equation}
\label{corot}
\rm R_{\rm co} = \left( \displaystyle \frac{\G M_{\star} }{\Omega_{\star}^2}  \right)^{1/3} = \left( \displaystyle \frac{\G M_{\star} \rm P_{\rm rot,\star}^2 }{(2\pi)^2}  \right)^{1/3},
\end{equation}
where $\rm P_{\rm rot,\star}$ is the surface rotation period of the host star, $\G$ the gravitational constant, $\rm M_{\star}$ the stellar mass, and $\Omega_{\star} = 2\pi / \rm P_{\rm rot,\star}$ the surface angular velocity of the host star.

Following \citet{GB15}, the evolution of the angular momentum rate of the convective envelope is given by
\begin{eqnarray} \label{dJconvdt}
  \displaystyle \frac{\dd \Jconv}{\dd t} = \Gamma_{\rm wind} + \Gamma_{\rm tide} + \Gamma_{\rm c-e} + \Gamma_{\rm rad_{evol}},
\end{eqnarray}
where  $\Gamma_{\rm wind}$,  $\Gamma_{\rm tide}$, $\Gamma_{\rm c-e}$, and $\Gamma_{\rm rad,evol}$ are the torques that act on the convective envelope;
$\Gamma_{\rm c-e}$ and $\Gamma_{\rm rad,evol}$ are torques applied on the convective envelope by the radiative core.
Figure~\ref{schema} shows the different torques exerted on the convective envelope of a low-mass star. For each of these components we use the following prescriptions.

\subsubsection*{Wind torque $\Gamma_{\rm wind}$}
\label{saturated}

The wind braking torque is given by \citep{Schatzman62,WD67}
\begin{eqnarray} 
\label{torque_wind}
\Gamma_{\rm wind} & \propto K_1^2 \Os \dot{M}_{\rm wind} r^2_A.
\end{eqnarray}
It depends on the stellar rotation rate $\Os$, mass-loss rate $\dot{M}_{\rm wind}$, and Alfv\'en radius $r_A$.
We use here the prescription of \citet{Matt12} for the Alfv\'en radius and the revised prescription of \citet{Cranmer11} for the mass-loss rate (see \citealt{GB15} for details). In our numerical code, $K_1$ is a free parameter and is associated to the wind braking efficiency. It is set so as to reproduce the observed rotation rate of the present Sun and rotational distribution of late-MS cluster.

The stellar wind braking $\Gamma_{\rm wind}$ relies on two parameters: the mass-loss rate and the value of the Alfv\'enic radius $r_A, $  which both depend on the evolution of the mean stellar magnetic field {$B_{\star} f_{\star}$ \citep[see][]{Cranmer11,Matt12}, where $B_{\star}$ is the stellar magnetic field strength and $f_{\star}$ is the magnetic filling factor (i.e. the fraction of the surface of the star that is magnetized). As in \citet{GB15}, we follow \citet{Cranmer11} by assuming that the magnetic field of the star is at the thermal equilibrium with the stellar photosphere and can thus be expressed as a function of the equipartition magnetic field strength, and $f_{\star}$ is a unique function of the Rossby number $\rm Ro = P_{rot,\star}/\tau_{\rm conv}$, that is, the ratio of the rotation period $\rm P_{rot,\star}$ to the convective turnover timescale $\tau_{\rm conv}$. Finally, the mean magnetic field can be expressed as}
\begin{eqnarray} 
 B_{\star} f_{\star} = 1.13\sqrt{\frac{8\pi\rho_{\star}  k_B  \rm T_{eff}}{\mu \rm m_{\rm H}}}   \frac{0.55}{\left[   1+(x/0.16)^{2.3}    \right]^{1.22}},
\end{eqnarray}
where $\rm x = Ro/Ro_{\odot}$ and $\rm Ro_{\odot} = 1.96$, $\rho_{\star}$ is the photospheric mass density, $k_B$ the Boltzmann's constant, $\rm T_{eff}$ the effective temperature, $\mu$ the mean atomic weight, and $\rm m_H$ the mass of a hydrogen atom \citep[see Eq. 3 from][]{GB15}.

Observations (e.g. X-ray luminosity; \citealt{Pizzolato03}, and magnetic flux density; \citealt{Reiners10}) show that the evolution of the magnetic field of low-mass stars reaches a plateau (hereafter the saturation or saturated regime) at a maximum value \citep[see Fig. 6 of][]{Reiners09} when the rotation of the star exceeds a certain threshold ($\approx 16~\Omega_{\odot}$ during the MS for a 1.0 $\rm M_{\odot}$ star).

Above this limit the mass-loss rate and magnetic field strength of fast rotating stars become constant regardless of the evolution of the surface rotation rate. In the saturated regime, the extraction of angular momentum through the stellar wind scales as $\Gamma_{\rm wind} \propto \Omega_{\star}$ \citep[since $B_{\star}$ and $\dot{M}_{\rm wind}$ become constant, see Eq. (\ref{torque_wind}) and Fig. 4 from][]{GB15} instead of scaling as $\Omega_{\star}^3$ (i.e. the empirical \citet{Sku72} relationship).

\subsubsection*{Core-envelope coupling torque $\Gamma_{\rm c-e}$}

As in \citet{MacGregorBrenner1991} we assume that both the core and the envelope are in solid body rotation with two different rotation rates and that a quantity $\Delta J$ of angular momentum is transferred between the core and the envelope over a timescale $\tau_{\rm c-e}$. The torque $\Gamma_{\rm c-e}$ associated to the angular momentum transfer rate between the radiative core and the convective envelope is expressed as
\begin{eqnarray}
\label{gammadeltaj}
\Gamma_{\rm c-e} = \displaystyle \frac{\Delta J}{\tau_{\rm c-e}},
\end{eqnarray}
with $\Delta J$ the quantity of angular momentum to be transferred, instantaneously, between the two regions to obtain a uniform rotation \citep{MacGregorBrenner1991},
\begin{eqnarray}
\label{DeltaJ}
\Delta J &=& \displaystyle \frac{\Iconv\Jrad-\Irad\Jconv}{\Irad+\Iconv}, \\
\label{DeltaJ2}
 &=& \displaystyle \frac{\Iconv\Irad}{\Irad+\Iconv} \left( \Osrad -\Osconv \right).
\end{eqnarray}
The coupling timescale $\tau_{\rm c-e}$ is a free parameter of our numerical code that is determined using the observed rotational period distribution of early-MS stellar clusters \citep[see][]{GB15}. It is related to internal transport mechanisms \citep[e.g.][and references therein]{Maeder09,Mathis13}

\subsubsection*{Radiative core torque $\Gamma_{\rm rad,evol}$}

Along the growth of the radiative core during the PMS phase, part of the convective envelope becomes radiative. Thus, we can consider that angular momentum is extracted from the envelope and transferred to the expanding radiative core. The associated torque $\Gamma_{\rm rad,evol}$ is \citep{Allain98}
\begin{eqnarray}
\label{gammarad}
\Gamma_{\rm rad,evol}= \displaystyle \frac{2}{3} R_{\rm rad}^2 \Osconv \displaystyle \frac{dM_{\rm rad}}{dt},
\end{eqnarray}
with $\rm M_{\rm rad}$ the mass of the radiative core. 

\subsection{Rotational evolution}

Finally, the evolution of the rotation rate of the convective envelope and that of the radiative core are given by 
\begin{eqnarray}
\label{domeg/dt}
\displaystyle \frac{\dd \Osconv}{\dd t} =&\displaystyle \frac{1}{\Iconv}\left(\Gamma_{\rm wind}+\Gamma_{\rm tide}+\Gamma_{\rm c-e}+\Gamma_{\rm rad,evol}\right) - \displaystyle \frac{\Osconv}{\Iconv} \displaystyle \frac{\dd \Iconv}{\dd t},  \\
\displaystyle \frac{\dd \Osrad}{\dd t} =&\displaystyle \frac{1}{\Irad}\left(\Gamma_{\rm rad,evol} - \Gamma_{\rm c-e}\right) - \displaystyle \frac{\Osrad}{\Irad} \displaystyle \frac{\dd \Irad}{\dd t}.
\end{eqnarray}
The rotation rate of the convective envelope can be decomposed into five contributions: the contraction, the stellar wind, the tides, the core-envelope coupling, and the development of the radiative core. The rotation rate of the radiative core is controlled by only three terms: the contraction, the development of the radiative core, and the transport of angular momentum between the core and the envelope. We therefore expect the rotational evolution of the envelope to be quite complex.
\begin{table}
\caption{Reference case.}
\begin{center}
\begin{tabular}{c|c}
Parameter               & value \\
\hline
\hline
$\Ms (\Msun)$           & 1.0   \\
$\rm P_{\star, \rm init}$ (day)    & 3.0   \\
$t_{\rm init}$ (Myr) & 5.0   \\
$\tau_{\rm c-e}$ (Myr)   & 0.5   \\
$\Mp (\Mjup)$           & 1.0     \\
\end{tabular} 
\end{center}
\label{ref_case}
\end{table}

Each run for a given star-planet configuration is thus characterized by a number of initial parameters. For the star they are the initial rotation rate of the star $\rm P_{\rm {rot, init}}$, the coupling timescale between the radiative core and the convective envelope $\tau_{\rm c-e}$, the disk lifetime $\tau_{\rm disk}$, the wind braking efficiency $K_1$, and the stellar mass $\rm M_{\star}$. For the planet they are the initial semi-major axis $a$ (hereafter SMA), the planetary mass $\rm M_{\rm p}$, and the initial time at which we consider the presence of the planet $t_{\rm init}$.

Since we start our simulations at the end of the disk lifetime, $t_{\rm init} \equiv \tau_{\rm disk}$ and corresponds to the age at which the tidal interaction starts. The initial SMA is also taken as a fraction of $R_{\rm co}$ and is thus a function of the initial stellar rotation rate $\rm P_{\rm rot, init}$. There is a degeneracy between $t_{\rm init}  \equiv \tau_{\rm disk}$ and $\rm P_{\rm rot, init}$ since using a longer $t_{\rm init}$ for a given  $\rm P_{\rm rot, init}$ can be similar (in terms of rotational evolution) to a short $t_{\rm init}$ but with a small $\rm P_{\rm rot, init}$. However, each $t_{\rm init}$ is associated to only one stellar internal structure and tidal dissipation efficiency couple. 

In the rest of this work, we adopt the parametrization extracted from \citet{GB15} that links the mass and initial rotation rate to the free parameters of our numerical code:
\begin{eqnarray}
\label{equaGB151}
\tau_{\rm c-e} &=& \rm P_{\rm{rot,init}}^{0.66} (\rm{days}) \times 10^{2.06}  \times 0.06 \left(\displaystyle \frac{M_{\star}}{M_{\odot}}\right)^{-3.97} \rm{Myr},  \\
\label{equaGB152}
\tau_{\rm disk} &=& \rm P_{\rm{rot,init}}^{0.86} (\rm{days}) \times 10^{0.17} \times \left(\displaystyle \frac{M_{\star}}{M_{\odot}}\right)^{1.55} \rm{Myr},  \\
\label{equaGB153}
\rm K_1 &=& 26.6 + 22.6\left(\rm \displaystyle \frac{M_{\star}}{M_{\odot}}\right)^2-47.5\displaystyle \frac{\rm M_{\star}}{\rm M_{\odot}}. 
\end{eqnarray}

\begin{figure*}[!ht]
    \begin{center}
        \includegraphics[width=17cm]{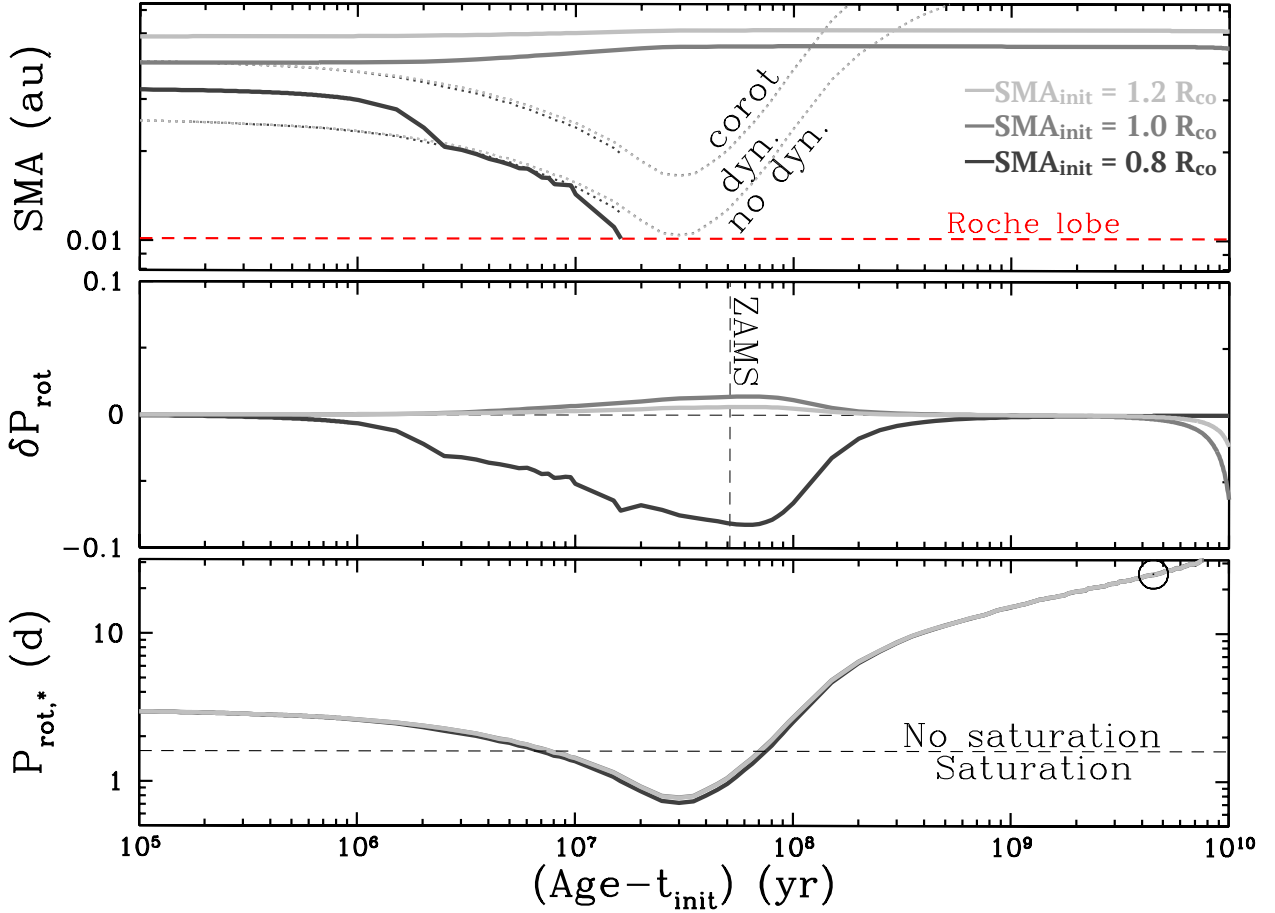} 
        \caption{Evolution of the orbital distance of a Jupiter-mass planet with different initial SMA (top panel) ranging from $0.8~R_{\rm{co}}$ (black), 1.0 $R_{\rm{co}}$ (dark grey), and  $1.2~R_{\rm{co}}$ (light grey), $\delta \rm P_{\rm rot}$ (middle panel), and the stellar rotation period (bottom panel) during the evolution of a 1.0 $\rm M_{\odot}$ star with an initial rotation period of three days and a coupling timescale $\tau_{\rm c-e}$ of 0.5 Myr. {Upper panel}: The orbital distance of the planet is represented in full coloured lines. The upper dotted lines correspond to $\rm P_{\rm orb}$ = $\rm \rm P_{\rm rot,\star}$ (the corotation limit) and the lower dotted lines correspond to $\rm P_{\rm orb}$ = 1/2 $\rm \rm P_{\rm rot,\star}$ limit, which marks the region above which the dynamical tide operates. The red long-dashed line is the evolution of the Roche lobe radius. {Middle panel}:  Departure of the rotation rate of the host star ($\rm \rm P_{\rm rot,\star}$) from an isolated star ($\rm \rm P_{\rm rot,isol.}$). The horizontal dashed line indicates no difference with an isolated star, and the vertical dashed line represents the localisation of the ZAMS (around 50 Myr for a 1.0 $\rm M_{\odot}$ star).  {Lower panel}: Rotation period (in days) of the host star. The horizontal dashed line correspond to the transition between saturated and unsaturated magnetic regimes (1.6 days $\approx 16~\Omega_{\odot}$), and the circle represents the rotation rate of the present Sun. 
The time on the x-axis is given from an initial time $t_{\rm init}$ of 5 Myr.}
    \label{Evolreferencecase}
 \end{center}
\end{figure*}

\section{Rotational evolution and star-planet systems}
\label{rotevol}

\subsection{Case of a solar mass star}
\label{referencecase}

The reference case used for comparison is given in Table \ref{ref_case}. We consider a one solar-mass star, with a rotation period of three days at 5 Myr \citep[between a moderate and a fast rotation rate regarding the observed rotation period distribution of early PMS clusters, see][]{ONC}, and a core-envelope coupling timescale of 0.5 Myr (this short timescale is chosen to mimic a solid body rotation where both convective and radiative regions have the same rotation rate). We consider a close-in Jupiter-mass planet and investigate the angular velocity of the star under the action of planetary tidal migration.
In this work we focus on initial SMA close to and within the corotation radius (see Eq. (\ref{corot})), since type I and type II migrations should bring the planets up to the corotation in the accretion disk on a short timescale \citep[0.01-0.1 Myr, see][]{Masset06}. 

\subsubsection{Generalities about planetary orbital evolution}

As described in Sect. \ref{tidaldissip}, there are two components of the stellar tides: equilibrium and dynamical. While the equilibrium tide is always present during the whole stellar life, the dynamical tide is only triggered when $\rm P_{\rm orb} > 1/2~\rm P_{\rm rot,\star}$ \citep{Bolmont16}. The orbital evolution of a given planet then depends on the value of the ratio $\rm \rm P_{\rm orb} / \rm \rm P_{\rm rot,\star}$, and in substance on which dominant tide the planet will be subject to during its evolution.
Moreover, for a given initial $\rm \rm P_{\rm orb} / \rm \rm P_{\rm rot,\star}$ configuration, a change of $t_{\rm init}$ leads to different stellar angular velocity and planetary orbital evolutions (since each $t_{\rm init}$ is associated to a single stellar internal structure-tidal dissipation properties couple).
In this work, the impact of the planetary orbital evolution on the surface rotation rate of the host star is estimated using the quantity $\delta \rm P_{\rm rot} = 1 - \Omega_{\rm conv,\star}/\Omega_{\rm conv,isol.} = 1 - \rm \rm P_{\rm rot,isol.} / \rm \rm P_{\rm rot,\star}$; $\delta \rm \rm P_{\rm rot}$ corresponds to the departure of the rotational evolution of the planet host star ($\rm \rm P_{\rm rot,\star}$) compared to an isolated (i.e. without planet) star ($\rm \rm P_{\rm rot,isol.}$).

~\\
Figure \ref{Evolreferencecase} shows the evolution of the SMA, $\delta \rm \rm P_{\rm rot}$, and stellar rotation $\rm \rm P_{\rm rot,\star}$ for the reference model with different initial SMAs ranging from 0.8 $R_{\rm co}$ to 1.2 $R_{\rm co}$, with $R_{\rm co} = 4.06\times10^{-2}$ au at $t_{\rm init}$.
During the first Myr after $t_{\rm init}$, there is no planetary migration and the planet marginally affects the surface rotation rate of its host star. Outward and inward planetary migrations start to be observed at about 1 Myr after $t_{\rm init}$, which corresponds to the age (i.e. around 6 Myr) at which the dissipation of the dynamical tide inside the stellar convective envelope is maximum in regard to the stellar internal structure \citep[see Fig. 4 of][]{Gallet17b}. The corresponding effect on the surface rotation rate of the star is shown in the middle panel of Fig. \ref{Evolreferencecase}. Inward and outward migrations globally lead respectively to the acceleration ($\delta \rm \rm P_{\rm rot} < 0$) and deceleration ($\delta \rm \rm P_{\rm rot} > 0$) of the star compared to an isolated star. 
\begin{SCfigure*}
        \includegraphics[width=12cm]{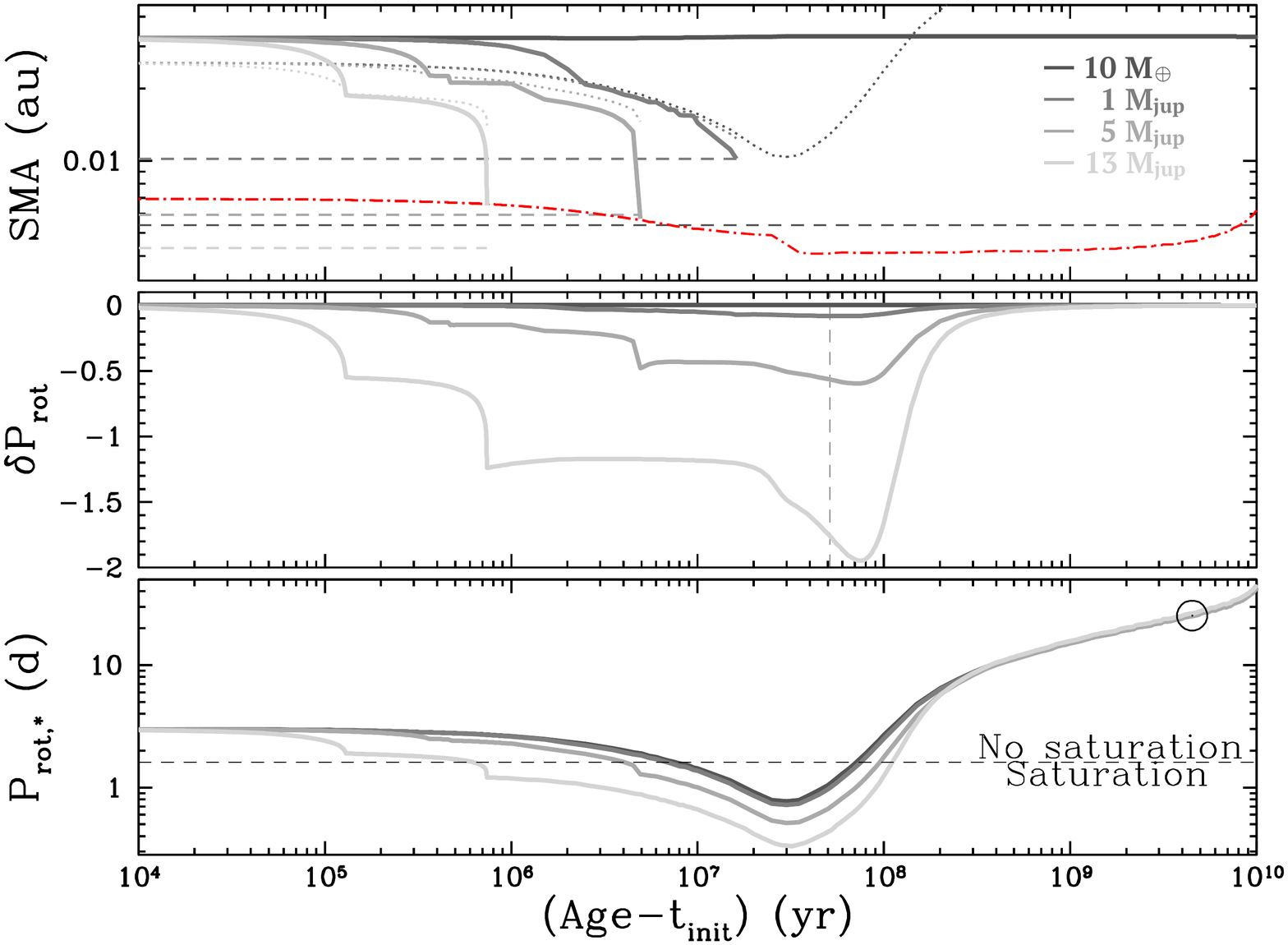} 
        \caption{Effect of planetary companion mass on the rotational evolution of a 1.0 $\rm M_{\odot}$ mass star. The initial rotation period of the star is three days, the star is assumed to rotate as a solid body with a core-envelope coupling timescale of 0.5 Myr; $\rm SMA_{init}$ = 0.8 $\rm R_{\rm co}\approx 3.25\times 10^{-2}$ au. The masses of the planets are 10 $\rm M_{\oplus}$, 1 $\rm M_{\rm Jup}$, 5 $\rm M_{\rm Jup}$, and 13 $\rm M_{\rm Jup}$, displayed as darkish to greyish coloured lines, respectively. {In the upper panel, the Roche lobe limits are displayed as long-dashed lines and the radius of the star as red dash-dotted line}. The horizontal dashed line in the lower panel corresponds to the transition between the saturated and unsaturated magnetic regime, and the circle represents the surface rotation rate of the present Sun. 
The time on the x-axis is given from an initial time $t_{\rm init}$ of 5 Myr.}
        \label{Evolmassplanet}%
\end{SCfigure*}

During the PMS phase, when the star is contracting, the surface rotation rate of the star is marginally impacted ($\delta \rm \rm P_{\rm rot}$ < -0.1) either by the planetary inward migration or by the planetary engulfment \citep[since $J_{\rm orb}/J_{\star} \approx 10^{-2}$, with $J_{\rm orb}$ the orbital angular momentum, see][]{Hut1981}.

\begin{SCfigure*}
        \includegraphics[angle=-90, width=12cm]{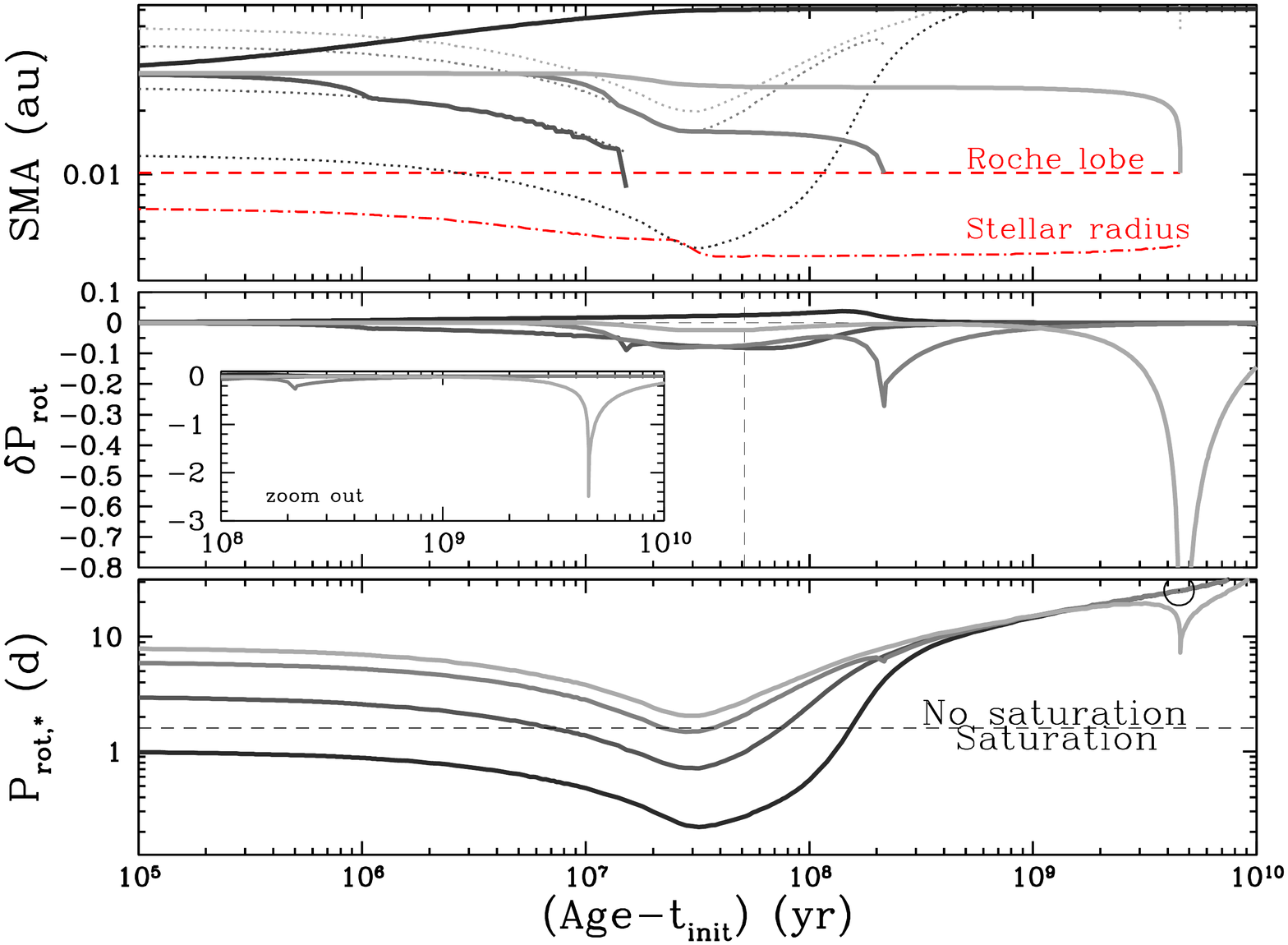} 
        \caption{Evolution of the orbital distance of a Jupiter-mass planet with an initial SMA of 0.03 au (top panel), $\delta \rm \rm P_{\rm rot}$ (middle panel), and the stellar rotation period (bottom panel) during the evolution of a 1.0 $\rm M_{\odot}$ star with four different initial rotation rates: one, three, six, and eight days from {darkish} to {greyish} coloured lines, respectively. In the upper panel, the dotted lines correspond to the $\rm \rm P_{\rm orb} = 1/2 \rm \rm P_{\rm rot,\star}$ limit, the red dashed line is the Roche lobe limit, and the red dashed-dotted line is the stellar radius.}
        \label{Evolrot}%
\end{SCfigure*}

On the MS, and because of the extraction of angular momentum by the stellar wind, the $\delta \rm \rm P_{\rm rot}$ of each configuration tends towards zero (i.e. the rotational convergence), which erases the knowledge of the presence of a massive planet from the rotational history of the star.

Finally, during the sub-giant phase and red-giant branch (hereafter RGB), all low-mass stars engulf their planet (because the planet either reaches the Roche lobe\footnote{The Roche limit is defined in \citet{Roche} and approximated by $(R_{\rm p}/0.462) * (M_{\rm p} /M_{\star})^{-1/3}$ au, with $R_{\rm p}$ the radius of the planet in au, and $\rm M_{\rm p}$ and $\rm M_{\star}$ the mass of the planet and the star, respectively, and traces the region in which the planet is considered to be engulfed by the star.} or the stellar radius). The surface rotation rate of the host stars is then strongly affected by this planetary engulfment leading to the rapid decrease of $\delta \rm \rm P_{\rm rot}$ \citep[see][for more details about the planetary engulfment during the RGB phase]{Privitera2016,Privitera2016b}.

\subsubsection{Exploration of the star-planet system parameters}
The path followed by the surface rotation rate of the star impacts both the orbital evolution of the planet (through its impact on the corotation radius and on the dynamical and equilibrium tide limit) while it also has an impact on the difference $\delta \rm \rm P_{\rm rot}$ of the planet host's star, compared to an isolated star, through the impact of orbital migration on the rotation of the star. Rotational evolution has thus a major role in star-planet interaction processes. In the following we describe the impact of the star-planet system's physical quantities on the rotational evolution of the star and its feedback effect on the orbital evolution of massive planets.

\subsubsection*{Impact of core-envelope decoupling timescale and initial time}
In the literature \citep[e.g.][]{Bolmont12,Bolmont16} the study of the orbital evolution of massive planet and its effect on the stellar surface rotation rate is often treated using solid body rotation for the whole star (see the cases in Fig. \ref{Evolreferencecase} with a coupling timescale of 0.5 Myr). Increasing the decoupling between the core and the envelope of the star allows the convective envelope to be braked earlier during the PMS phase. As a consequence, the corotation radius increases (since $R_{\rm co} \propto \rm \rm P_{\rm rot, \star}^{2/3}$, see Eq. (\ref{corot})) and the tidal torque is reduced (since $\Gamma_{\rm tide} \propto |1 - \Os/n|$, see Eq. (\ref{torque_tid})). This allows the planet to survive longer, by 10-15 Myr compared to the solid body rotation case. {Indeed, in the case of the stellar reference star and with a 1 $\rm M_{jup}$ planet initially located at $\rm 0.8~R_{co}$ ($\approx$ 0.03 au), increasing the coupling timescale between the core and the envelope from 0.5 Myr (solid body) to 10 and 30 Myr \citep[that correspond to the parametrization of the fast and slow rotators, see][]{GB15} induces a shift in the planetary engulfment of $\approx$ 5 and 15 Myr, respectively.}

Similarly, decreasing $t_{\rm init} \equiv \tau_{\rm disk} $ induces an earlier spin-up of the star because of the contraction. This allows the planet to survive the stellar evolution due to a decrease in $R_{\rm co}$. {In the case of the stellar reference star and with a 1 $\rm M_{jup}$ planet initially located at $\rm 0.8~R_{co}$ ($\approx$ 0.03 au), increasing the disk lifetime from 2 to 5 Myr leads to the inward (engulfment) migration of the planet.}

\subsubsection*{Impact of planetary mass}

Figure \ref{Evolmassplanet} shows the implication of a change of the mass of the planet, ranging from 10 $\rm M_{\oplus}$ to 13 $\rm M_{\rm Jup}$, on the planetary orbital evolution and stellar rotational evolution. Here we assume the same configuration as in the reference case, except for the planetary mass, and we consider an initial SMA of 0.8 $\rm R_{\rm co}\approx 3.25\times 10^{-2}$ au. The effect of the planetary mass is indeed quite strong on both SMA and stellar surface rotation rate evolution.

For a given star-planet configuration, when the mass of the planet increases, the migration timescale decreases causing the planet to migrate earlier during the stellar evolution (see Eq. \ref{Ts}). The mass of the planet also affects the strength of the impact of the planetary migration on the surface rotation evolution of the host star. Higher mass means higher orbital angular momentum. At the Zero age main-sequence (hereafter ZAMS) this leads to $\delta \rm \rm P_{\rm rot}$=-1.5 -- -3 for the 13 $\rm M_{\rm Jup}$ and to $\delta \rm \rm P_{\rm rot}$=-0.5 for the 5 $\rm M_{\rm Jup}$. For planets less massive than 1 $\rm M_{\rm jup}$, no significant effect on the surface rotation of the star is seen up to the end of the MS phase, consistent with previous theoretical works \citep{Bolmont16} and observations \citep{Ceillier16}.

\begin{figure*}[!ht]
    \begin{center}
        \subfigure[Initial SMA (in \%$R_{\rm co}$)]{
            \label{percentsub}
            \includegraphics[width=0.45\linewidth]{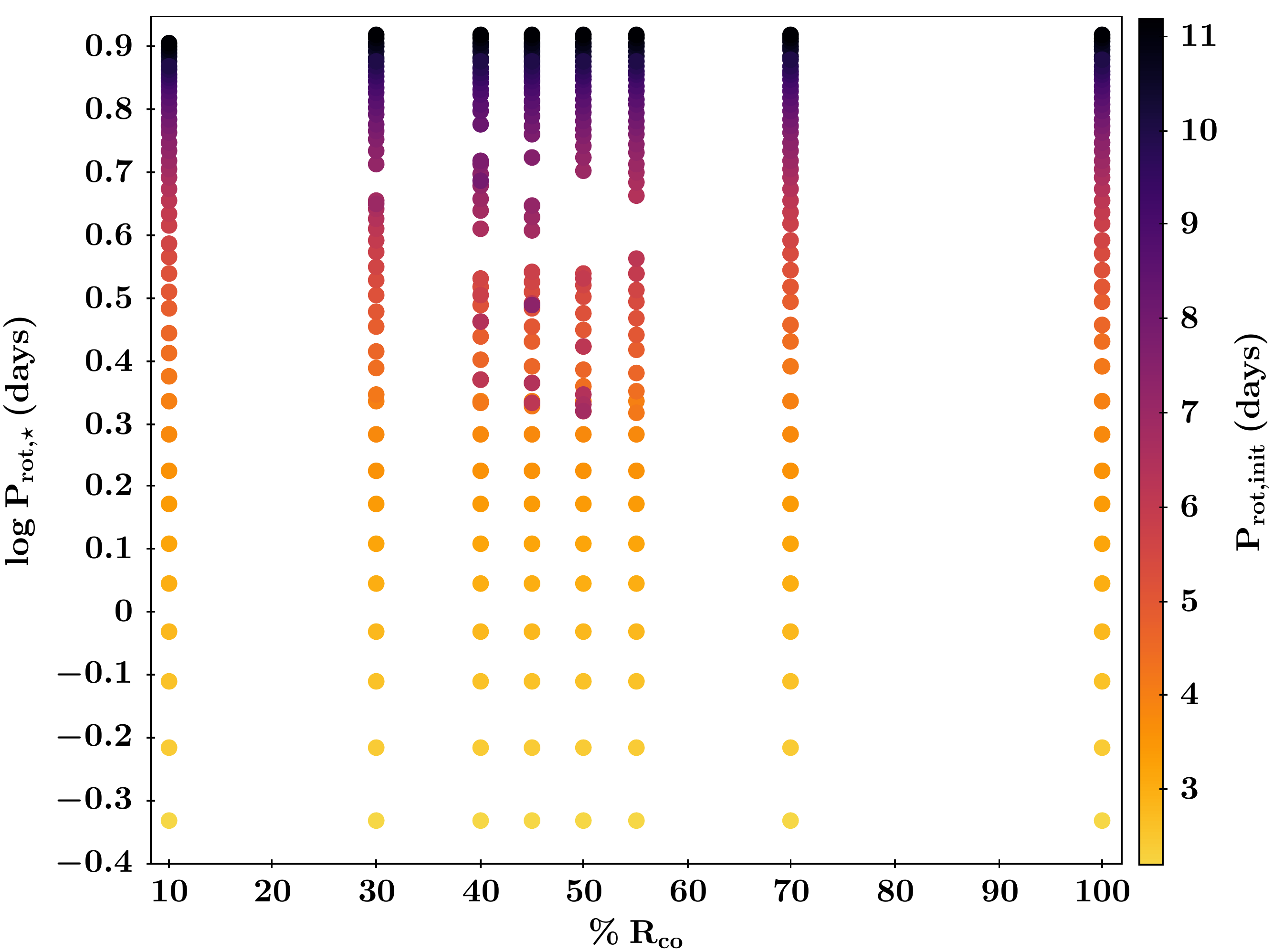} 
        } 
       \hspace{-0.1cm}
        \subfigure[Mass planet]{
           \label{massplanetsub}
           \includegraphics[width=0.45\linewidth]{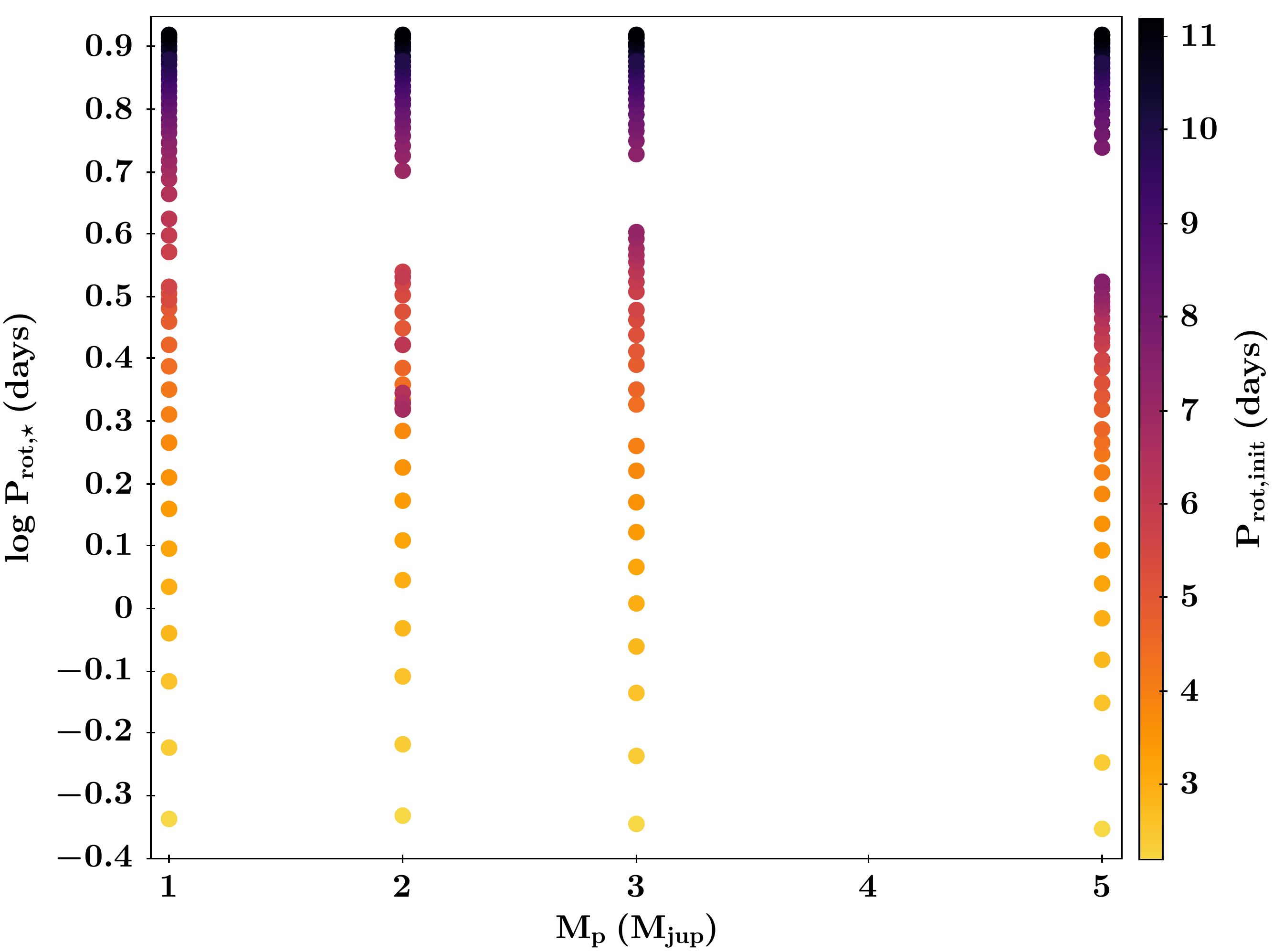}
        } 
        \caption{Distribution at 120 Myr in log $\rm \rm P_{\rm rot,\star}$ obtained for a 0.8 $\rm M_{\odot}$ from an initial distribution of rotation period ranging from two to 11 days. Effect of the initial SMA (10, 30, 45, 50, 55, 70, and 100\%  $\rm R_{\rm co}$) in the case of a 2 $\rm M_{\rm jup}$ (\ref{percentsub}). Impact of the mass of a planet (between 1 and 5 $\rm M_{\rm jup}$) initially located at 50\%  $\rm R_{\rm co}$ (\ref{massplanetsub}). These distributions are extracted at 120 Myr that is about the age of the Pleiades cluster.}
        \label{percent}%
    \end{center}
\end{figure*}

\subsubsection*{Initial SMA inside the corotation radius}

Finally, we fix the initial SMA of the planet at 0.03 au and explore the impact of the initial rotation rate of the star using one, three, six, and eight days. Figure \ref{Evolrot} shows the evolution of the planetary SMA and stellar surface rotation rate for a Jupiter-mass planet orbiting a 1.0 $\rm M_{\odot}$ mass star. For each initial rotation period, the initial SMA corresponds to a fraction of the corotation radius. With an initial SMA of 0.03 au, we have
\begin{itemize}
        \item One day: $\rm SMA_{init}$ = 1.53 $\rm R_{\rm co}$,
        \item Three days: $\rm SMA_{init}$ = 0.73 $\rm R_{\rm co}$,
        \item Six days: $\rm SMA_{init}$ = 0.46 $\rm R_{\rm co}$,
        \item Eight days: $\rm SMA_{init}$ = 0.38 $\rm R_{\rm co}$.
\end{itemize}

In Fig. \ref{Evolrot}, three different cases can be described. The first one is the case where the planet migrates outward already during the PMS. This happens because the initial rotation period of the star is sufficiently fast (one day) so that the initial value of the SMA (0.03 au) is located outside of the corotation radius. Because the planet is initially outside of the corotation radius and its evolution is determined by the dynamical tide, it is submitted to an efficient outward torque.

The second case is the one where the planet falls onto the stellar surface during the PMS phase. It appears when the initial rotation rate of the star is three days. In that configuration, the planet is initially located between  $\rm  1/2~\rm \rm P_{\rm rot,\star} < \rm \rm P_{orb} < \rm \rm P_{\rm rot,\star}$ and is thus dominated by the dynamical tides. In that case, the Jupiter-mass planet falls into the star during the PMS phase and does not significantly affect its rotation rate.

The last case is the one where the planet falls into the stellar surface during the MS phase. This happens when the initial rotation rate of the star is low (six to eight days). With such low initial rotation rate the planet is initially inside of the $\rm  1/2~\rm \rm P_{\rm rot,\star}$ limit and is thus only subject to the equilibrium tide. Because the equilibrium tide is less efficient than the dynamical tide, the planet migrates inward on much longer timescales (of the order of $10^9$ years) and thus falls onto the stellar surface much later. For this case, the impact of the planet engulfment on the surface rotation rate of the star is quite important as it reaches a spin rate four times larger than that of an isolated star.

\begin{figure*}
    \begin{center}
        \subfigure[Mass planet = 1 $\rm M_{\rm jup}$]{
            \label{Distribtime1mjup}
            \includegraphics[angle=-90,width=0.47\linewidth]{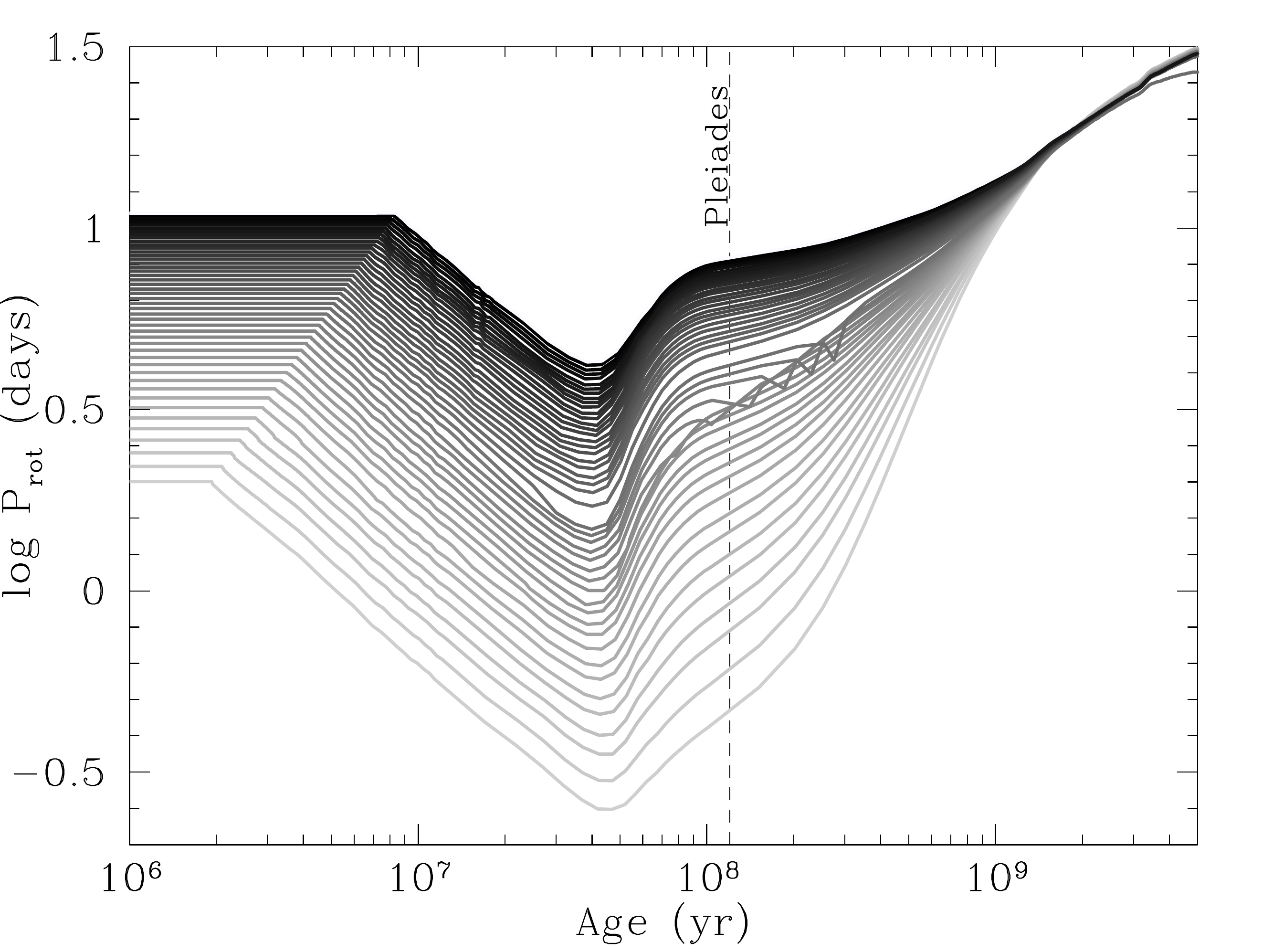} 
        } 
        \hspace{-0.5 cm}
        \subfigure[Mass planet = 2 $\rm M_{\rm jup}$]{
           \label{Distribtime2mjup}
           \includegraphics[angle=-90,width=0.47\linewidth]{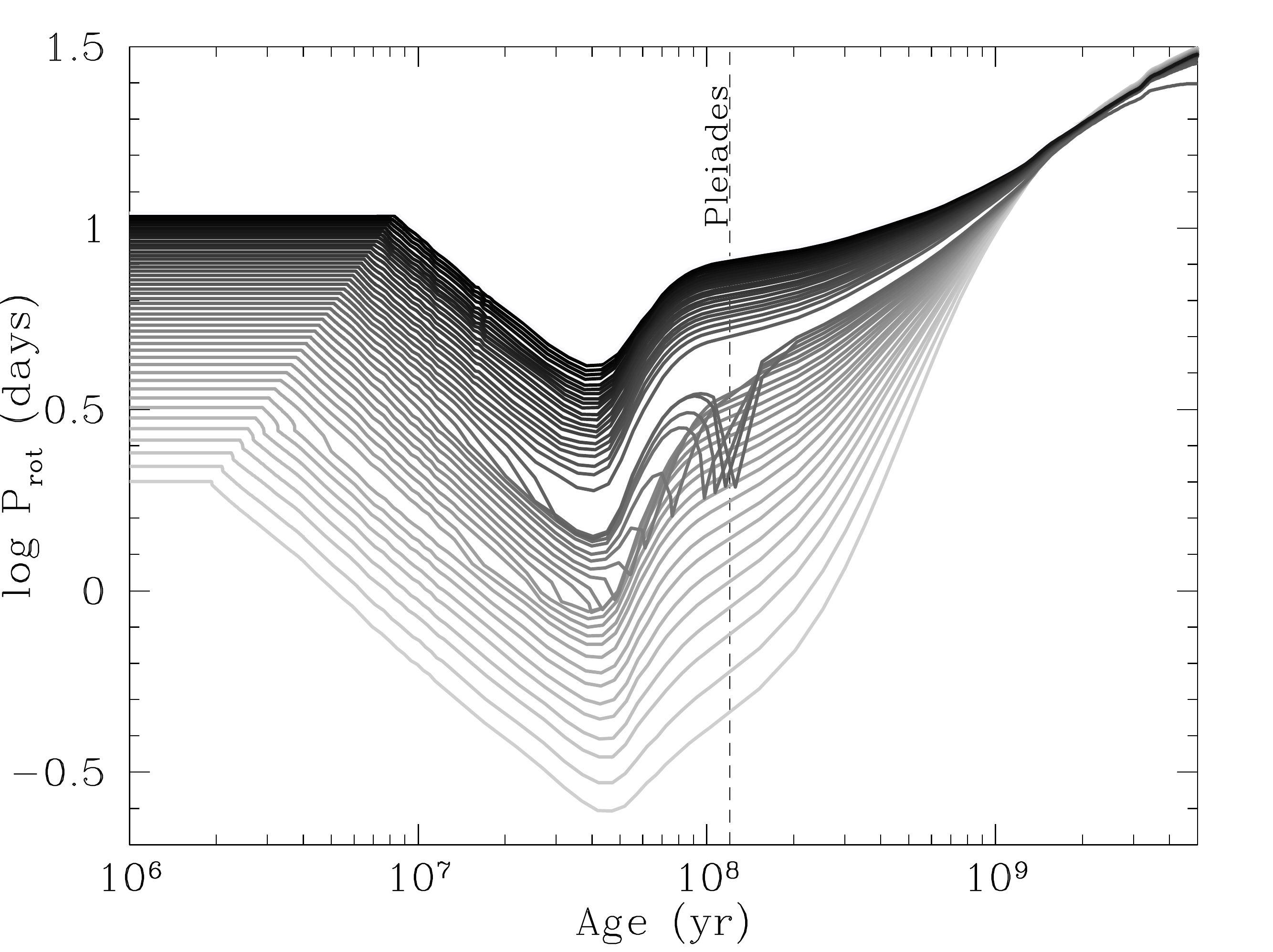}
        }
         \caption{Temporal evolution of the rotation distribution related to the 0.8 $\rm M_{\odot}$ mass star. The mass of the planet is taken to be 1 $\rm M_{\rm jup}$ (\ref{Distribtime1mjup}) and 2 $\rm M_{\rm jup}$ (\ref{Distribtime2mjup}) and is initially located at 0.5 $R_{\rm co}$. Each line correspond to one stellar configuration. The bottom {grey} line corresponds to an initial period of two days, the upper {black} line corresponds to an initial period of 11 days. The consecutive difference between each line is 0.2 day.}
        \label{Distribtime}%
    \end{center}
\end{figure*}
\subsection{Evolution of star-planet initial distributions}
\label{distribinit}

The most important parameters that are involved in the star-planet rotational-orbital evolution are the initial location SMA$_{\rm init}$ of the planet around the host star (expressed in $R_{\rm co}$) and the mass of the planet $\rm M_{\rm p}$. To analyse the impact of the SMA$_{\rm init}$-$\rm M_{\rm p}$ space, we first evolve in time an initially uniform (in terms of surface rotation rate) distribution consisting of 0.8 $\rm M_{\odot}$ solar metallicity stars with an initial rotation period ranging from one to 11 days (with a 0.2 days step). 

\subsubsection{Corotation radius}

\begin{SCfigure*} 
        \includegraphics[width=12cm]{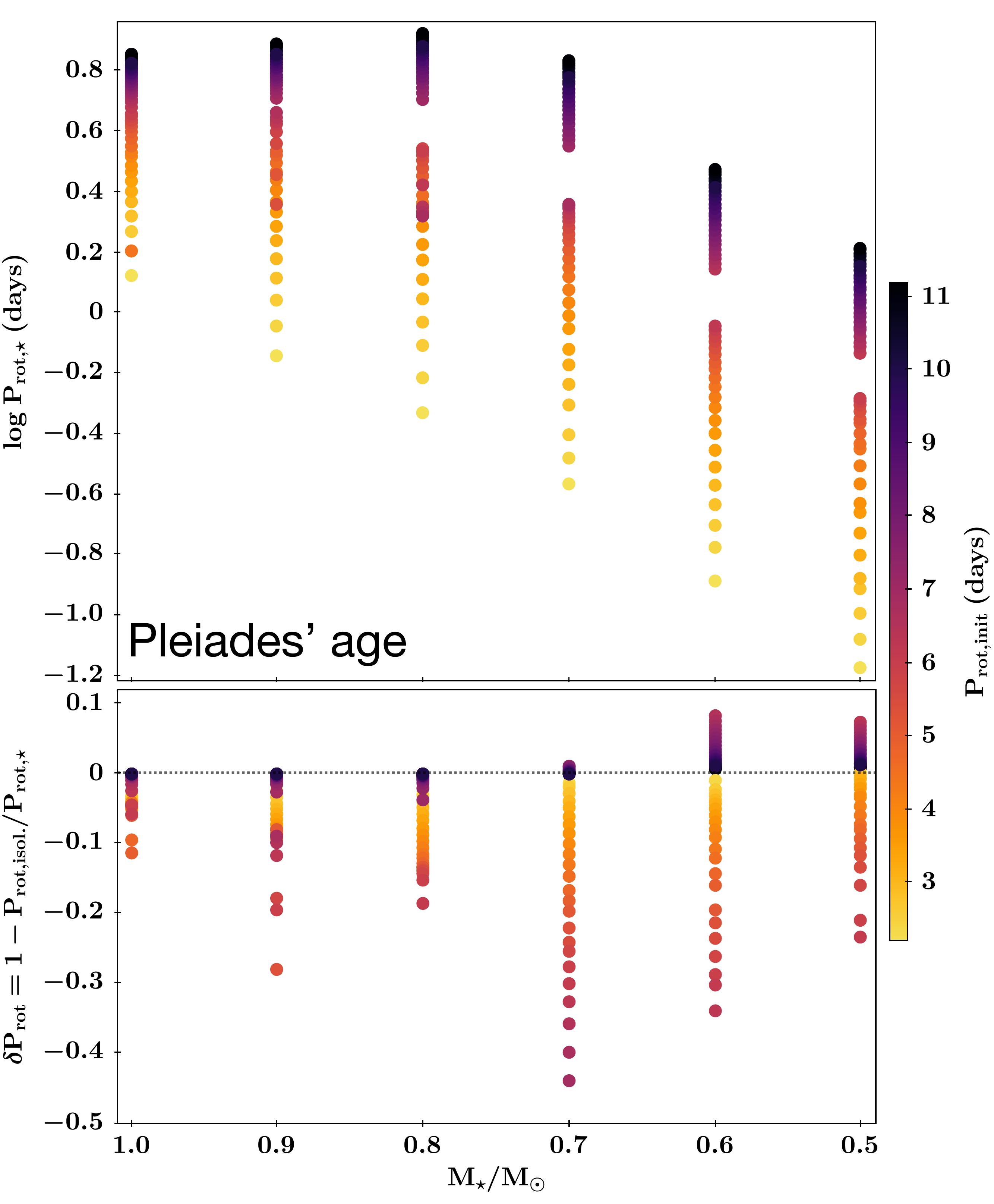} 
        \caption{{Upper panel:} Rotation period distribution of stars between 0.5 and 1.0 $\rm M_{\odot}$ at 120 Myr. The mass of the planet is taken to be 2 $\rm M_{\rm jup}$. Each coloured plain circle corresponds to one stellar configuration. The initial SMA of the planet is $\rm 50\%~R_{\rm co}$. {Lower panel: Corresponding values of $\delta \rm P_{rot}$ for each of the star-planet systems as a function of the stellar mass.}}
        \label{Distrib}%
\end{SCfigure*}

The greatest impacts of tidal interactions on surface rotation rate are located, for a 0.7-0.8 $\rm M_{\odot}$ mass star, in a narrow range of initial rotation period around five to eight days. We consider here that around each star orbits a 2 $\rm M_{\rm jup}$ mass planet whose initial SMA is taken to be 10, 30, 40, 45, 50, 55, 70, and 100\% $\rm R_{\rm co}$. Figure \ref{percentsub} shows each of these distributions for a 0.8 $\rm M_{\odot}$ star at 120 Myr (the age of the Pleiades) as a function of the initial localization of the planet. During the MS phase, the rotational period distributions are the most impacted by the planet when it is located around 50 $\pm$ 5\% of the corotation radius.
 
In the case of an initial SMA of 50\%  $\rm R_{\rm co}$, the planets are initially inside the equilibrium tide regime. Hence, they remain on the same orbit during the PMS phase since equilibrium tide is less efficient than the dynamical tide. However, while the rotation rate of the star increases (due to the stellar contraction), the corotation radius slowly moves close to the stellar surface. At some point during the end of the PMS phase, the planet crosses the limit 1/2 $\rm \rm P_{\rm rot,\star}$ and becomes sensitive to the dynamical tide and thus starts to migrate faster. The planet then falls into the stellar surface during the early-MS phase, and strongly impacts its surface rotation rate.

At 30\% or 70\% $\rm R_{\rm co}$, the planet falls during the PMS which, as pointed out above, has almost no effect on the MS stellar rotational evolution. For the first case, it is because the planets are initially too close to the star and thus either fall into the stellar surface during the PMS phase, or during the MS phase at older ages than the age of the Pleiades, hence not creating any kink in the rotation period distribution at 120 Myr. For 70\% $\rm R_{\rm co}$, the most close-in planets fall into the stellar surface during the PMS phase, the others migrate outward. The limit $ \rm \rm P_{orb} = 1/2~\rm \rm P_{\rm rot,\star}$ corresponds to $2^{-2/3}R_{\rm co} \approx 63\%~R_{\rm co}$.

\subsubsection{Massive planet}
We also explored the impact of the planetary mass on the opening of the rotational kink. Figure \ref{Distribtime} shows the temporal evolution of the rotational distribution related to the 0.8 $\rm M_{\odot}$ mass star in the case of a 1 $\rm M_{\rm jup}$ (Fig. \ref{Distribtime1mjup}) and 2 $\rm M_{\rm jup}$ (Fig. \ref{Distribtime2mjup}). In these figures we can see a {rotational kink} that starts to be opened between 20-30 Myr up to the early MS phase in each case. The size of the {rotational kink} is more important when the planetary mass increases. The amplitude of the {rotational kink} is at its maximum during the early MS phase and is slowly reduced by the rotational convergence due to the magnetic braking \citep[see][]{GB15}.

The tidal star-planet interaction is at its maximum for $\rm SMA_{init}= 0.5~R_{\rm co}$. In the case of a 0.8 $\rm M_{\odot}$ star, only the intermediate rotators ($\rm \rm P_{\rm rot,init}$ = 5-8 days) are impacted by the tidal interactions, which implies that not all planets in our initial rotational distribution are involved in the production of the {rotational kink}.

\subsection{Exploration of stellar mass}
\label{stellarmass}
We explore a larger range of stellar mass, namely 0.5-1.0 $\rm M_{\odot}$, and consider a 2 $\rm M_{\rm jup}$ planet so as to increase the {rotational kink} in $\rm log~\rm \rm P_{\rm rot}$ observed in Fig. \ref{percent} around $\rm log~\rm \rm P_{\rm rot} = 0.5-0.6$. 
We then evolve the same initial distribution as described above for initial planetary SMA around 50\% $\rm R_{\rm co}$. 
Table \ref{simus} shows part of the parametrization used for each of the stars from the star-planet configurations explored in this article. This table summarizes the value of the coupling timescale $\tau_{\rm c-e}$ (Myr) and initial time $t_{\rm init}$ (Myr), as well as the value of the wind braking constant $K_1$ for each stellar mass and non-decimal rotation period from one to 11 days. These values are estimated using Eqs. (\ref{equaGB151}-\ref{equaGB153}) and allow the \citet{GB15} models to reproduce the observed rotational distribution of low-mass stars. In our simulations, the rotation period step is 0.2 days.

Figure \ref{Distrib} shows the rotation period distribution of stars between 0.5 and 1.0 $\rm M_{\odot}$ at 120 Myr. We consider here that around each star orbits a 2 $\rm M_{\rm jup}$ mass planet initially located at $\rm 50\%~R_{\rm co}$. The colour gradient corresponds to the value of the initial rotational period of the star (between two and 11 days). The presence of a planet more massive than 1 $\rm M_{\rm jup}$ creates a kink in the rotational period distribution of stars with mass smaller than 0.9 $\rm M_{\odot}$. The location of the {rotational kink} depends on the stellar mass (it is around log $\rm \rm P_{rot,\star}=0.6$, corresponding to $\rm \rm P_{rot,\star}=4$ days, for 0.8 $\rm M_{\odot}$ stars, and around log $\rm \rm P_{rot,\star}=0$, corresponding to $\rm \rm P_{rot,\star}=1$ days, for the 0.6 $\rm M_{\odot}$ stars). However, at the age of the Pleiades cluster, the tidal star-planet interactions only affect stars with initial rotational period of around seven days (the greenish plain circle). Moreover, this {rotational kink} appears by the end of the PMS phase and persists (depending on the planetary mass, see Fig. \ref{Distribtime}) up to the early MS phase. 

\begin{table*}[ht]
\caption{Parametrization of the modelled stars as a function of their mass and initial rotation period. The listed value are $\tau_{\rm c-e}/t_{\rm init}$ expressed in Myr and the last column gives the value of the wind braking constant $K_1$ that only depends on the stellar mass. The rotational periods are expressed in days. We only show in this table the non-decimal part of the rotation period distribution. In our simulations, the rotation period step is 0.2 days.} 
\label{simus}
\begin{small}
\begin{tabular}{|c|c|c|c|c|c|c|c|c|c|c|c|c|}
\hline 
\backslashbox{Mass}{$\rm \rm P_{\rm rot,init}$} & 1 & 2 & 3 & 4 & 5 & 6 & 7 & 8 & 9 & 10 & 11 & $K_1$  \\
\hline 
\hline 
0.5 $\rm M_{\odot}$     &       115/0.5 &       182/0.9 &       238/1.3 &       288/1.7 &       334/2.0 &       376/2.4 &       417/2.7 &       455/3.1 &       492/3.4 &       527/3.7 &       561/4.0 & 8.5 \\
0.6 $\rm M_{\odot}$     &       56/0.7          &       88/1.2          &       115/1.7 &       140/2.2 &       162/2.7 &       182/3.2 &       202/3.6 &       221/4.1 &       238/4.5 &       256/4.9 &       272/5.4 & 6.2\\
0.7 $\rm M_{\odot}$     &       30/0.9          &       48/1.6          &       63/2.2          &       76/2.8          &       88/3.5          &       99/4.0          &       110/4.6 &       120/5.2 &       129/5.7 &       139/6.3 &       148/6.8 & 4.4\\
0.8 $\rm M_{\odot}$     &       18/1.1          &       28/1.9          &       37/2.7          &       45/3.5          &       52/4.2          &       58/5.0          &       64/5.7          &       70/6.4          &       76/7.1          &       82/7.7          &       87/8.4 & 3.1\\
0.9 $\rm M_{\odot}$     &       11/1.3          &       18/2.3          &       23/3.3          &       28/4.2          &       32/5.1          &       36/6.0          &       40/6.8          &       44/7.6          &       48/8.5          &       51/9.3          &       54/10.1 & 2.2\\
1.0 $\rm M_{\odot}$     &       7/1.5           &       12/2.7          &       15/3.9          &       18/4.9          &       21/6.0          &       24/7.0          &       27/8.0          &       29/9.0          &       31/10.0 &       34/10.9 &       36/11.9 & 1.7\\
\hline 
\end{tabular} 
\end{small}
\end{table*}

{{We recall that this article does not aim to reproduce the observed rotational distribution as this has been previously investigated \citep[e.g.][]{GB13,GB15,SA17} but rather to explore a large star-planet parameter space, in particular the initial rotational period.}} { {In this work, we do not consider the very external slow rotation part of the distributions but more likely the first and third percentiles \citep[see][]{GB13,GB15} of these distributions. As a consequence, we do not reproduce the outliers (at least the very slow rotating stars) of these distributions at any ages. Given our uniform initial rotational distribution that ranges between one and 11 days, it was expected that our model should miss parts of the observations' properties.}}

{ {There is also a deviation in the fit of $\rm \tau_{disk}$ that produces a smaller disk lifetime than inferred from \citet{GB15} for the low-mass stars of our sample. As a consequence, the whole rotational distribution of 0.6 and 0.5 $\rm M_{\odot}$ stars is shifted towards a small rotational period.}}

\subsection{Comparison with the Pleiades observations}
\label{explomass}
In the previous sections, we pointed out that the only way for the surface rotation rate of the star to be significantly affected by the star-planet interaction is 1) when the planet is initially located inside the corotation radius (around 50\% of $R_{\rm co}$), and 2) when the planet is more massive than 1 $\rm M_{\rm jup}$. We thus decided to explore the parameter space of these two main parameters to try to reproduce the \citet{Rebull16} and \citet{Stauffer16} observations and the so-called rotational anomalies described below. We would like to point out that in this work we do not aim to reproduce all of the features observed in the Pleiades cluster (e.g. the wide rotational distribution of the less massive stars) but only the ``kink'' around 0.8 $\rm M_{\odot}$ stars. 
\begin{figure}[!h]
    \begin{center}
        \includegraphics[angle=0, width=0.9\linewidth]{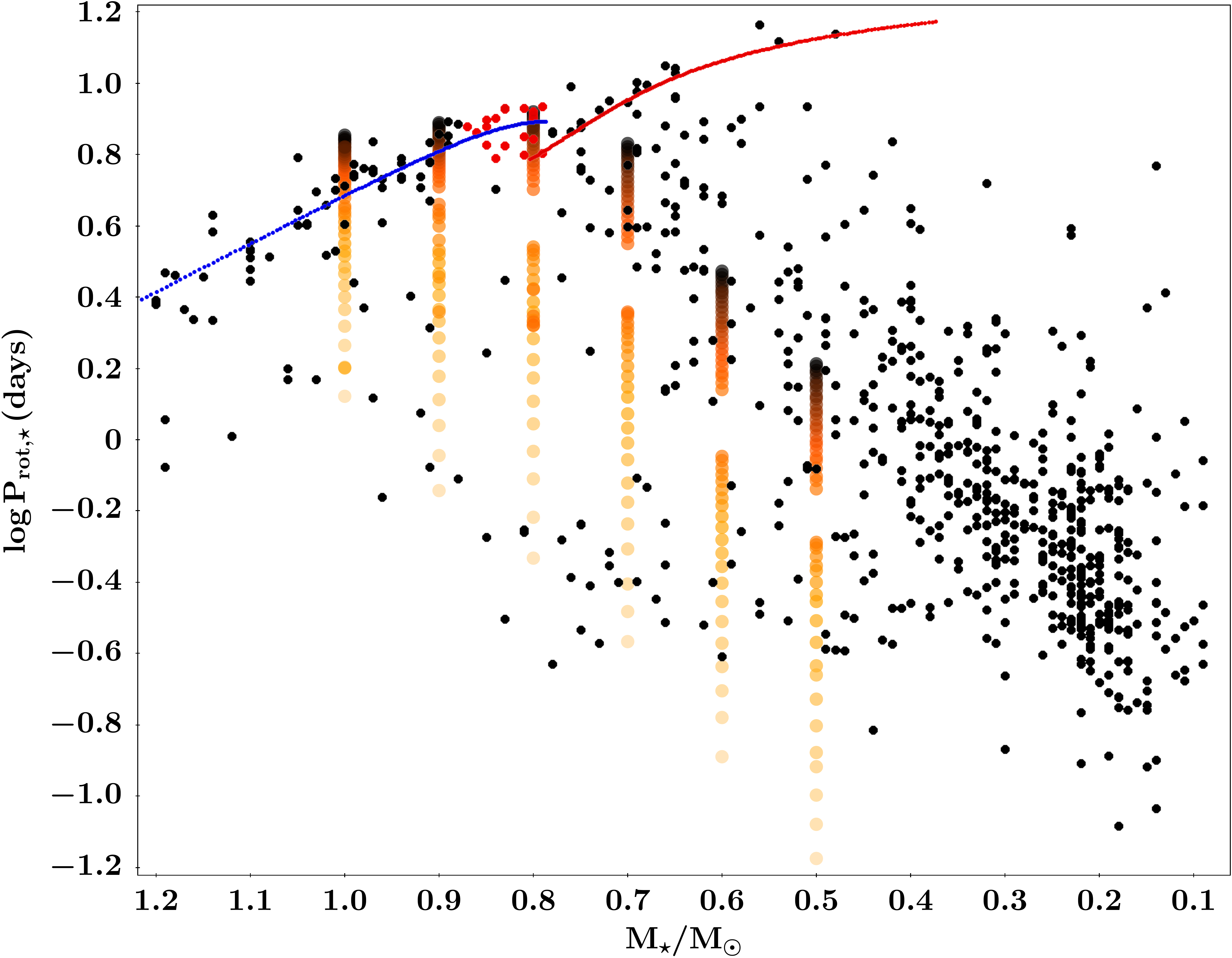}  
        \caption{Rotation period distribution of stars between 0.1 and 1.2 $\rm M_{\odot}$ from the 120 Myr Pleiades cluster. The red circles correspond to the stars that display the rotational kink. Data are from \citet{Rebull16} and \citet{Stauffer16}. The stellar masses are estimated using the absolute $K_s$ magnitudes and the 125 Myr isochrones from \citet{Baraffe15}. The blue and red lines correspond to the least-squares fit to the slowest rotating stars inferred by \citet[][see their Fig. 2]{Stauffer16}. The rotational distributions are the same as in Fig. \ref{Distrib}.}
        \label{Stauffer}%
    \end{center}
\end{figure}

Figure \ref{Stauffer} shows the rotation period distribution of stars between 0.1 and 1.05 $\rm M_{\odot}$ from the 120 Myr Pleiades cluster (also known as M45 and Melotte 22). In this figure we can see the ``anomaly'' detected by \citet{Rebull16} and \citet{Stauffer16} in the Pleiades cluster obtained using {Kepler-K2} \citep{K2}. It is depicted by a {rotational kink} located around {slow rotating (five to eight days)} 0.8 $\rm M_{\odot}$ stars (early K-type stars). In Fig. \ref{Stauffer} it is highlighted by the red plain circles {in which part of the slowly rotating stars appear to rotate with a faster rotation than the ``classical'' slow sequence}. In this figure, there are two populations: low- ($\rm M_{\star} /M_{\odot} \lesssim 0.5 $) and intermediate- ($0.5 \lesssim M_{\star}/M_{\odot}  \lesssim 1.1$) mass stars. The low-mass stars display a quite large rotational distribution mainly located at small rotational period ($\rm P_{rot,\star} \approx 0.3$ days, corresponding to $\rm log~\rm P_{rot,\star}=-0.52$). Since these stars are largely (or totally if  $\rm M_{\star} /M_{\odot} < 0.3$) convective, they are less efficiently braked by the stellar wind and thus are still on the fast rotating regime at the age of the Pleiades cluster. In contrast, the intermediate-mass stars harbour a more packed distribution centred around a rotational period that depends on the stellar mass (for instance the 0.8 $\rm M_{\odot}$ stars peak around eight days, $\rm log~\rm P_{rot,\star}=0.9$). Because these stars are partly convective, they are more efficiently braked by the stellar wind and thus possess a more compact distribution even at as early an age as 120 Myr. Among the slowest rotating stars a scatter can be seen. This anomalous scatter could be created by a physical process that should accelerate part of the slow rotators at the age of the Pleiades cluster. 

According to \citet{Stauffer16}, this anomaly is statistically robust and suggests two distinct regimes, one producing the slowest rotating stars that follows the classical rotational evolution, which only invokes magnetic braking and internal redistribution of angular momentum and the other that requires another external physical effect to produce faster rotators at the same mass and age. It is observed around the early K-type 0.8 $\rm M_{\odot}$ (V-K$_s \sim 2.6$) stars for which the dissipation of the dynamical tides is maximum at ZAMS \citep{Mathis15,Gallet17b}. Moreover, it seems to be time dependent as it appears to evolve toward lower masses with increasing age \citep[see][]{Stauffer16}. 
This kink should then be time and mass dependent and should be present during the early MS phase (around 120-220 Myr) and for early K-type stars (around 0.7-0.8 $\rm M_{\odot}$). Can tidal star-planet interactions reproduce these observations? If it is the case, then what could be the insights about the initial distribution of massive close-in planets around low-mass PMS stars, such as the initial location and the mass of the planets so as to reproduce the observations in the Pleiades?

Our model predicts that this {rotational kink} should also be present for slightly less massive stars, namely 0.7-0.6 $\rm M_{\odot}$, corresponding to $V-K_{s} \approx 3.3$. However, this behaviour is not observed in the data from \citet{Stauffer16}. It could not be explained by a shift in the mass estimation between the STAREVOL code used in this work and the \citet{Baraffe16} code used in \citet[][]{Stauffer16} since both models are quite similar and isochrones from these two models produce the same theoretical stellar masses (Louis Amard's PhD thesis and private communication). However, given the quite large scatter in the rotational distribution of 0.5-0.6 $\rm M_{\odot}$ stars, it is arduous to distinguish the presence or not of a given {rotational kink}.

{{In this work, we arbitrarily assumed an occurrence of a massive close-in planet of 100\% so as to illustrate the impact of the effect of a tidal star-planet interaction on the rotational evolution of low-mass stars. The net result of this interaction on the stellar rotational distributions naturally depends on the initial occurrence of massive planets around young low-mass stars. To investigate this, more realistic (i.e. non-uniform) initial stellar rotational and planetary distributions are first of all required. Such a complete study is, however, beyond the scope of the present paper. }}

We demonstrate that the presence of a massive planet can have a notable impact on the rotation period distribution of ZAMS's stellar clusters, especially for the K-type stars. While the modelled {rotational kink} is produced in the correct stellar mass range, the observed {rotational kink} in the Pleiades is located around slower rotating stars (five to eight days) than predicted by our model. It means that a physical process impacting preferentially very slow initial rotating stars is required to only impact the rotational period distribution of this stellar population. In this work, we only consider the tidal dissipation inside the convective envelope of the host star. {Including the additional dissipation by gravity waves inside the radiative core could move the {rotational kink} towards a longer rotational period. Indeed, the planets that are engulfed during the early-MS phase would be translated outwards compared to their current location (with the present model) and would then modify the rotation rate of slower rotating MS stars (because of the magnetic braking)}. This additional dissipation should then be included in future works, so as to produce more realistic star-planet evolutions.

\section{Conclusions}
\label{conclusion}

We investigated the theoretical effect of tidal interaction on the rotation rate of stars by coupling the parametric model described in \citet{GB15}, which includes the decoupling between the core and the envelope, to the planetary orbital evolution code of \citet{Bolmont16}. We find that the stellar rotation is primarily impacted during planetary engulfment events. 

With this work we showed that if the planets fall into the stellar surface during the early-MS phase, then planetary accretion has a strong effect on the surface rotation rate of the star. This is especially true in the case of a massive close-in planet (one to two Jupiter-mass planet) orbiting a low-mass star. The surface rotation rate of the star can then be accelerated up to a factor of two to three. Within the right configuration, a planetary population between 1 and 5 $\rm M_{\rm jup}$ initially located around 50\% $R_{\rm co}$ can open a {rotational kink} during the early MS phase, at the age of the Pleiades cluster. This {rotational kink} is only present for stars with initial surface rotation rate around seven days and located around $\rm log~\rm P_{\rm rot,\star}=0.6$ (corresponding to $\rm P_{\rm rot,\star}= 4$ days) and stellar mass smaller than 0.8 $\rm M_{\odot}$. {In \citet{Stauffer16} this {rotational kink} is however observed around $\rm log~\rm P_{\rm rot,\star}$= 0.8 (corresponding to  $\rm P_{\rm rot,\star}$= 6.3 days)}.

The proposed model here thus does not allow us to exactly reproduce the location of the observed anomaly, possibly because we neglected the tidal dissipation in the stellar radiative core. However, it provides a promising direction to further investigate the influence of close-in planets on the rotational period distribution of young stars, beyond the well-known period-mass relationships, especially in the framework of the Gaia DR2 \citep{Lanzafame18} and future PLAnetary Transits and Oscillations of stars (PLATO)/Transiting Exoplanet Survey Satellite (TESS) missions \citep{Plato,TESS}.

\begin{acknowledgements}
{We thank the anonymous referee and the editor for their fruitful comments and suggestions. We thank our colleague Philippe Delorme for the very useful discussions of some aspects of this paper.} F.G acknowledges financial support from the CNES fellowship. The authors acknowledge financial support from the Swiss National Science Foundation (FNS) and from the French Programme National de Physique Stellaire PNPS of CNRS/INSU.  F.G. and C.C. acknowledge financial support from the Swiss National Science Foundation and 294 SEFRI for project C.140049 under COST Action TD 1308 Origins (PI C.C.). This work results from the collaboration of the COST Action TD 1308. S.M. and E.B. acknowledge funding by the European Research Council through ERC grant SPIRE 647383 and support from the CNES-PLATO grant at CEA Saclay. This project has received funding from the European Research Council (ERC) under the European Union's Horizon 2020 research and innovation programme (grant agreement No 742095; {\it SPIDI}: Star-Planets-Inner Disk-Interactions).
\end{acknowledgements}

\bibliographystyle{aa} 
\bibliography{references} 

\end{document}